\documentclass[nofootinbib,aps,prd,preprint,superscriptaddress]{revtex4-1}
\usepackage{amsmath,graphicx,hyperref}

\newcommand{\fms}[1]{{#1}\!\!\!/}

\newcommand{\mc}{\mathcal}
\newcommand{\mr}{\mathrm}

\newcommand{\mO}{\mathcal{O}}

\newcommand{\be}{\begin{equation}} 
\newcommand{\ee}{\end{equation}} 
\newcommand{\bea}{\begin{eqnarray}} 
\newcommand{\eea}{\end{eqnarray}} 

\newcommand{\ov}{\overline}
\newcommand{\pp}{\perp}
\newcommand{\dg}{\dagger}
\newcommand{\n}{\overline{n}}
\newcommand{\nn}{\frac{\fms{\overline{n}}}{2}} 
\newcommand{\nnn}{\frac{\fms{n}}{2}} 

\newcommand{\bl}[1]{{\bf{#1}}}

\newcommand{\blp}[1]{{\bf{#1}}_{\perp}}
\newcommand{\blpu}[1]{{\bf{#1}}^{\perp}}

\newcommand{\bsp}[1]{{\boldsymbol{#1}}_{\perp}}

\newcommand{\nnb}{\nonumber} 

\newcommand{\as}{\alpha_s}

\newcommand{\veps}{\varepsilon} 

\newcommand{\UV}{\veps_{\mr{UV}}}
\newcommand{\IR}{\veps_{\mr{IR}}}

\newcommand{\La}{\Lambda^2_{\rm{alg}}}

\begin{document}

\baselineskip 3.0ex 

\vspace*{18pt}


\title{Fragmentation of a Jet with Small Radius}

\def\Seoultech{Institute of Convergence Fundamental Studies and School of Liberal Arts, Seoul National University of Science and Technology, Seoul 01811, Korea}
\def\Pitt{Pittsburgh Particle Physics Astrophysics and Cosmology Center (PITT PACC) \\ Department of Physics and Astronomy, University of Pittsburgh, Pittsburgh, Pennsylvania 15260, USA}

\author{Lin Dai}
\email[E-mail:]{lid33@pitt.edu}
\affiliation{\Pitt}
\author{Chul Kim}
\email[E-mail:]{chul@seoultech.ac.kr}
\affiliation{\Seoultech} 
\author{Adam K. Leibovich}
\email[E-mail:]{akl2@pitt.edu}
\affiliation{\Pitt}  

\begin{abstract} 
\baselineskip 3.0ex   
In this paper we consider the fragmentation of a parton into a jet  with small jet radius $R$. Perturbatively, logarithms of $R$ can appear, which for narrow jets can lead to large corrections.  Using soft-collinear effective theory, we introduce the fragmentation function to a jet (FFJ), which describes the fragmentation of a parton into a jet.  We discuss how these objects are related to the standard jet functions.  Calculating the FFJ to next-to-leading order, we show that these objects satisfy the standard Dokshitzer-Gribov-Lipatov-Altarelli-Parisi evolution equations, with a natural scale that depends upon $R$.  By using the standard renormalization group evolution, we can therefore resum logarithms of $R$. We further use the soft-collinear effective theory to prove a factorization theorem where the FFJs naturally appear, for the fragmentation of a hadron within a jet with small $R$.  Finally, we also show how this formalism can be used to resum the ratio of jet radii for a subjet to be emitted from within a fat jet.  
\end{abstract}

\maketitle 


\section{Introduction} 

Jets in high energy collisions have been an important theoretical and experimental probe of physics for decades. Currently, they are not only  important for understanding Quantum Chromodynamics (QCD), but are crucial in our searches for beyond the Standard Model physics at the Large Hadron Collider at CERN and will continue to be important for any future collider that may be built.  Understanding the property of jets and being able to calculate reliable cross sections to compare to data are thus extremely important to current and future studies in particle physics.  

Conceptually, a jet is a collinear set of energetic particles in the detector.  In order to make this concept concrete, there needs to be some jet algorithm to define how particles are sorted to be within or outside of the jet.  Most jet algorithms use a parameter to differentiate the two sets of particles, often denoted as the jet radius $R$. When doing theoretical calculations involving a jet  definition, logarithms of this new object occur, $\ln R$, and thus to make sure that we have perturbative convergence of QCD, choosing $R\sim 1$ would be natural.  However, it is sometime useful to investigate narrow jets by choosing a smaller $R$, since it can help resolve individual jets, remove pileup, and  probe jet substructure.  This leads to the problem of the breakdown of perturbation theory and requires the resummation of $\ln R$.  This has been investigated in QCD in Refs.~\cite{Dasgupta:2014yra,Dasgupta:2016bnd,Jager:2004jh,Mukherjee:2012uz}. Since jets are made up of collinear particles, soft-collinear effective theory (SCET) \cite{Bauer:2000ew,Bauer:2000yr,Bauer:2001yt,Bauer:2002nz} is a natural tool to study jets.  Indeed, there have been many studies of $\ln R$ resummation within SCET \cite{Becher:2015hka,Chien:2015cka,Becher:2016mmh,Kolodrubetz:2016dzb}.

In this paper, we introduce the fragmentation function to a jet (FFJ) in SCET.  The FFJ, $D_{J_k/l}(z,\mu)$, describes the fragmentation of parton $l$ into a jet with momentum fraction $z$ containing parton $k$. We calculate the different possible combinations of quark and gluon initial and final partons.  By summing over final state partons, we obtain the inclusive FFJ, $D_{J/l}(z,\mu)$, describing the inclusive fragmentation of parton $l$ into a jet. The renormalization of this object will be shown to lead to the standard Dokshitzer-Gribov-Lipatov-Altarelli-Parisi (DGLAP) evolution with the natural scale dependent on $R$, and thus we can use this object and the renormalization group to resum logarithms of $R$.  

We also present a new factorization theorem for the fragmentation of a hadron within a jet, where the FFJ appears, allowing for the resummation of $\ln R$ for this process.  We further generalize this factorization for the situation of a subjet with radius $r$ within a fat jet of radius $R$.  This allows the resummation of the ratio of these radii, $\ln R/r$.

The organization of this paper is as follows.  In sec.~\ref{incjff}, we give the definition of the FFJ in SCET, and calculate next-to-leading (NLO) corrections.  From this we can derive the renormalization group behavior and see that it is the standard DGLAP evolution.  In sec.~\ref{Facincjet} we present the factorization theorem for the fragmentation inside a jet. Combining this with the renormalization behavior from the previous section allows for resummation of  $\ln R$ for this process. In sec.~\ref{subjet} we consider the subjet fragmentation from a fat jet.  We conclude in sec.~\ref{conclude}.  Finally, in appendix~\ref{A1} we describe hadron fragmentation inside of a jet, which is very similar to the subjet fragmentation of sec.~\ref{subjet}.

Please note that while completing this work, Ref.~\cite{Kang:2016mcy} appeared on the arXiv with significant overlap with our sec.~\ref{incjff}.

\section{Inclusive Jet Fragmentation Function} 
\label{incjff}

The definition of the fragmentation function to a jet (FFJ) is similar to the fragmentation function to a hadron (HFF).  In SCET, if a collinear quark, $q$, fragments to a jet with a momentum fraction $z$, the probability is given as 
\bea 
D_{J_k/q}(z,\mu) = \sum_{X_{\notin J},X_{J-1}}\frac{1}{2N_cz}  \int d^{D-2}\blpu{p}_J &&  \mr{Tr} \langle 0 | \delta \Bigl(\frac{p_J^+}{z}-\mc{P}_+\Bigr)\delta^{(D-2)}(\bsp{\mc{P}})\nn \Psi_n | J_k(p_J^+, \blpu{p}_J, R) X_{\notin J}\rangle \nnb\\ 
\label{JFFparf}
~~~~~~&&\times\langle J_k(p_J^+, \blpu{p}_J, R) X_{\notin J} | \bar{\Psi}_n |0\rangle,
\eea
where $\Psi_n=W_n^{\dagger} \xi_n$, $W_n$ is a collinear Wilson line in SCET~\cite{Bauer:2000yr,Bauer:2001yt}, and $R$ is the jet radius to be determined by specific jet algorithm. $X_{J-1}$ are the final states included in the observed jet except the primary jet parton $k$ and $X_{\notin J}$ are final states not included in the jet. Throughout, we will work in $D=4-2\veps$ dimensions, and use the convention, $p_+\equiv \n\cdot p = p_0 + \hat{\bl{n}}_J\cdot \bl{p}$, $p_-\equiv n\cdot p = p_0 -\hat{\bl{n}}_J\cdot \bl{p}$, where $\hat{\bl{n}}_J$ is an unit vector in the jet direction. The lightcone vectors $n$ and $\n$ satisfy $n^2=\n^2=0$ and $n\cdot\n =2$. Therefore $p_+ \sim 2E$ for a collinear particle in $\hat{\bl{n}}_J$ direction. The expression of FFJ in Eq.~(\ref{JFFparf}) is displayed in the parton frame, where the transverse momentum of the mother parton, $\blp{p}$, is zero.

If we consider FFJ in the jet frame, where the transverse momentum of the observed jet, $\blpu{p}_J=0$, we can do the integral on $\blpu{p}_J$ using the relation $\blp{p} = - \blpu{p}_J/z$. As a result we can express FFJ  as 
\be \label{JFFjetf} 
D_{J_k/q}(z,\mu) = \sum_{X_{\notin J},X_{J-1}} \frac{z^{D-3}}{2N_c} \mr{Tr} \langle 0 | \delta \Bigl(\frac{p_J^+}{z}-\mc{P}_+\Bigr) \nn \Psi_n | J_k(p_J^+,R) X_{\notin J}\rangle 
\langle J_k(p_J^+,R) X_{\notin J} | \bar{\Psi}_n |0\rangle.
\ee
The normalization is chosen so that at lowest order (LO) in $\as$, the FFJ is given by
\be\label{JFFLO}
D_{J_q/q}^{(0)}(z) = \frac{z^{D-3}}{2N_c} \mr{Tr} \nn p_J^+ \nnn \delta \left(\frac{p_J^+}{z}-p_J^+\right)\cdot N_c = \delta(1-z). 
\ee
Like usual fragmentation functions to hadrons (HFFs), the FFJ satisfies the following momentum conservation, 
\be\label{mcon}
\sum_{k=q,\bar{q},g} \int^1_0 dz z D_{J_k/q} (z,\mu) = 1.
\ee

When a gluon initiates a jet fragmentation, the gluon FFJ in the parton frame is defined as 
\bea\label{gJFFpar}
D_{J_k/g} (z,\mu) &=&  \sum_{X_{\notin J},X_{J-1}} \frac{1}{p_J^+(D-2)(N_c^2-1)} \int d^{D-2}\blpu{p}_J  \\
&&\hspace{-1cm}\times \mr{Tr} \langle 0 |  \delta\Bigl(\frac{P_+}{z}-\mc{P}_+\Bigr)\delta^{(D-2)}(\bsp{\mc{P}})  \mc{B}_n^{\pp\mu,a}
| J_k(p_J^+,\blpu{p}_J,R) X_{\notin J}\rangle 
\langle J_k(p_J^+,\blpu{p}_J,R) X_{\notin J} | \mc{B}_{n\mu}^{\pp a}| 0 \rangle_. \nnb
\eea
Here $\mc{B}_{n}^{\pp a}$ is a covariant collinear gluon field strength, defined by $\mc{B}_{n}^{\pp\mu,a}  = i\n^{\rho}g_{\perp}^{\mu\nu} G_{n,\rho\nu}^b \mc{W}_n^{ba} = i\n^{\rho}g_{\perp}^{\mu\nu} \mc{W}_n^{\dagger,ba} G_{n,\rho\nu}^b$, where $\mc{W}_n$ is the collinear Wilson line in the adjoint representation. It satisfies   
\be\label{Bperp} 
\mc{B}_{n}^{\pp\mu} = \mc{B}_{n}^{\pp\mu,a} T^a = \frac{1}{g} W_n^{\dg} \Bigl[\n\cdot iD_n, iD_n^{\pp\mu}\Bigr]W_n = \frac{1}{g} \Bigl[\mc{P}_+W_n^{\dg} iD_n^{\pp\mu} W_n \Bigr]_.
\ee

For defining the jet, we will employ an inclusive $\mr{k_T}$-type algorithm. This is a recombinational algorithm, which has the same constraint for $\mr{k_T}$~\cite{Catani:1993hr,Ellis:1993tq}, C/A~\cite{Dokshitzer:1997in}, and anti-$\mr{k_T}$~\cite{Cacciari:2008gp} up to NLO in $\as$. 
If two particles merge into a jet, the constraint is given by 
\bea
\label{kTe+e-} 
&&\theta < R~~~(e^+e^-~\mr{collider}), \\
\label{kThad}
&&\theta < \frac{R}{\cosh y}~~~(\mr{hadron~collider}), 
\eea
where $\theta$ is the angle between two particles.
For a hadron collider, we assumed $\Delta y$ and $\Delta \phi$ are small, so Eq.~(\ref{kThad}) is applicable to the jet with small $R$. When we compute NLO corrections to the jet algorithm, 
we will use $\theta < R'$ for the sake of simplicity, where  $R'=R$ for $e^+e^-$ colliders and $R'=R/\cosh y$ for hadron colliders. As we will see later, typical scales for jet functions are $p_+ \tan (R'/2)$. In the small $R$ limit, $p_+ \tan (R'/2) \sim E R'$ are approximated as $ER$ for $e^+e^-$ annihilation and $p_T R$ for hadron collision, where $p_T$ is the transverse momentum of the jet to the hadron beam direction. 

\begin{figure}[t]
\begin{center}
\includegraphics[height=5cm]{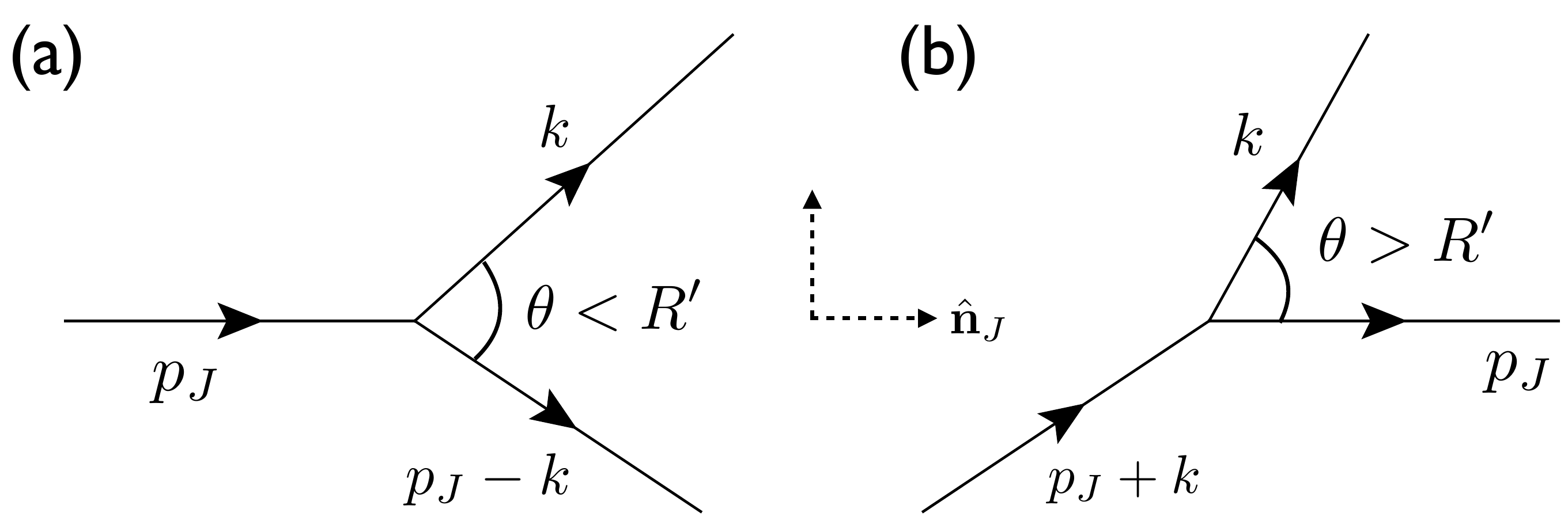}
\end{center}
\vspace{-0.3cm}
\caption{\label{fig1} \baselineskip 3.0ex 
Jet fragmentation at NLO in $\as$. Diagram (a) shows the jet merging, so the contribution to FFJ should be proportional to $\delta(1-z)$. Diagram (b) shows the jet splitting, which has a contribution with a fraction $z<1$. }
\end{figure}

Fig.~\ref{fig1} shows the two possible cases for jet fragmentation. 
If $\theta < R'$, shown in Fig.~\ref{fig1}-(a), the two particles in the final states are combined into a jet and the jet fraction is proportional to $\delta(1-z)$. In this case the phase space constraint in the jet frame ($\blpu{p}_J = 0$) is given by~\cite{Ellis:2010rwa}
\be\label{pscin} 
\tan^2 \frac{R'}{2} > \frac{p_J^{+2} k_-}{(p_J^+-k_+)^2 k_+}\ .
\ee
If $\theta>R'$, only one particle is chosen to be in the jet, shown in Fig.~\ref{fig1}-(b), hence the jet splitting arises with the fraction $z$. The phase space constraint in the jet frame becomes  
\be\label{pscout} 
\tan^2 \frac{R'}{2} < \frac{k_-}{k_+}\ . 
\ee

There appears to be a gap in the phase space between Eqs.~(\ref{pscin}) and (\ref{pscout}). 
However when we express the momentum of the mother parton as $p$, we have $\blp{p}=\blpu{p}_J=0$ for Eq.~(\ref{pscin}) but $\blp{p}=\blp{k}$ for Eq.~(\ref{pscout}) with $p_J=p-k$. Therefore when we express $k_-$ in terms of $p_+$ and $p^2$, $k_-$ is different in Eq.~(\ref{pscin}) and (\ref{pscout}); $k_-=(1-x)p^2/p_+$ for Eq.~(\ref{pscin}) and $k_-=p^2/((1-x)p_+)$ for Eq.~(\ref{pscout}), where $x = k_+/p_+$. So 
the right sides of the inequalities  
Eqs.~(\ref{pscin}) and (\ref{pscout}) end up both equaling $p^2/(x(1-x)p_+^2)$ and there is no gap in phase space.  

\subsection{NLO Calculation of Quark FFJ}
\label{subsec:qjffnlo}

Following the description in Fig.~\ref{fig1}, it is convenient to separate the full NLO contribution into `jet merging' ($\theta <R'$) and `jet splitting' ($\theta >R'$) contributions. In the jet merging contribution, the momentum of the mother parton is equal to the jet momentum, $p_J$. For quark initiated jets, it can be described by 
\bea\label{in} 
D^{\rm{in}}_{J/q}(z;E_JR') &=& \delta(1-z) \int^{\Lambda^2}_0 dM^2  \frac{1}{2N_c~p_J^+}  \\
&&~~~\times\sum_{X_{J-1}}\mr{Tr} \langle 0 | \delta (M^2 - \mc{P}^{2}) \nn \Psi_n | J_q (p_J^+,R)  \rangle \langle J_q (p_J^+,R) | \bar{\Psi}_n | 0 \rangle, \nnb
\eea
where $M^2$ is the invariant mass of the final states. The gluon case is similarly defined with $\mc{B}_{n}^{\pp\mu,a}$. $\Lambda^2$ is the maximal jet mass when $\theta = R'$. As there are two particles in the final state, $\Lambda^2$ is usually also dependent on each particle's energy. This jet merging contribution includes all the virtual corrections. Therefore combining the real and virtual contributions we can cancel all the infrared (IR) divergences and the result has only ultraviolet (UV) divergences.

Note that other than the $\delta(1-z)$, Eq.~(\ref{in}) is closely related to the standard quark jet function in SCET, defined as 
\be\label{jetf}
\sum_{X_n} \langle 0 | \Psi_n^{\alpha} | X_n \rangle \langle X_n | \bar{\Psi}_n^{\beta} | 0 \rangle 
= \int \frac{d^4 p_{X_n}}{(2\pi)^3} p_{X_n}^+ \nnn J_q (p_{X_n}^2) \delta^{\alpha\beta}. 
\ee
Here $J_q$ is normalized as $J_q^{(0)} (p^2)= \delta (p^2)$ at LO in $\as$. 
Using this, we can rewrite Eq.~(\ref{in}) to be
\be\label{in1} 
D^{\rm{in}}_{J/q}(z;E_JR')  = \delta(1-z) \int^{\Lambda^2}_0  dM^2 J_{q}(M^2;\theta<R') = \delta(1-z)\mc{J}_q (E_JR';\theta<R'), 
\ee
where $J_q(M^2;\theta<R')$ is the unintegrated jet function for the final states inside the jet and $\mc{J}_q$ is the integrated jet function (also called the unmeasured jet function in Ref~\cite{Ellis:2010rwa}). Both have been computed to NLO in Ref.~\cite{Ellis:2010rwa,Cheung:2009sg,Chay:2015ila} with $\mr{k_T}$-type and cone-type algorithms applied. When we apply the $\mr{k_T}$-type algorithm in Eq.~(\ref{kTe+e-}), the jet merging contribution to NLO is given by 
\bea\label{MinkT} 
D^{\rm{in}}_{J/q}(z;E_JR')&=& \delta(1-z)\Biggl\{1+\frac{\as C_F}{2\pi} \Biggl[\frac{1}{\UV^2}+\frac{1}{\UV}\Bigl(\frac{3}{2} +\ln\frac{\mu^2}{p_{J}^{+2}t^2}\Bigr) \\ 
&&~~~~~~~~~~~
+\frac{3}{2}\ln\frac{\mu^2}{p_{J}^{+2}t^2}+\frac{1}{2}\ln^2\frac{\mu^2}{p_{J}^{+2}t^2}+\frac{13}{2}-\frac{3\pi^2}{4} \Biggr]\Biggr\}\ ,\nnb
\eea 
where $t\equiv\tan(R'/2)\sim R'/2$.

Note that the renormalization behavior of the unintegrated jet function $J_{q}(M^2;\theta<R')$ in Eq.~(\ref{in1}) is different from the standard jet function without the restriction in Eq.~(\ref{jetf}). 
For example, all the UV divergences of the unintegrated jet function are only proportional to $\delta(M^2)$ while this is not true for the  standard jet function. 
The main reason for this difference comes from different treatments of the zero-bin subtraction~\cite{Manohar:2006nz}. For the unintegrated jet function in the small $R$ limit, the relevant zero-bin subtracted mode should  specifically be a collinear-soft mode~\cite{Becher:2015hka,Chien:2015cka,Bauer:2011uc} with scaling $(p_{cs}^+,p_{cs}^{\perp},p_{cs}^-) \sim Q\eta(1,R,R^2)$, where $\eta$ is a small parameter. This mode can resolve the jet boundary. Since the contribution of this collinear-soft mode to the jet mass squared is much smaller than $M^2 \sim E_J^2 R'^2$, UV divergences coming from this mode's zero-bin subtraction only contribute to  the $\delta(M^2)$  part. The details of the computation with this collinear-soft mode have been shown in Ref.~\cite{Chay:2015ila}. 
However, in case of the standard jet function, the zero-bin subtracted mode is an ordinary soft mode and its contribution to the jet mass is non-negligible. For this type of  zero-bin subtraction we  obtain UV divergences proportional to $1/M^2$ as well as $\delta(M^2)$.

In addition, there have been some complications about the integrability relation between the unintegrated and the integrated jet functions in Eq.~(\ref{in1}).
When $M^2 \sim E_J^2 R'^2$ in the small $R$ limit as considered in this paper, we can describe the unintegrated jet function using only the collinear mode scaling as $(p_+,p_{\pp},p_-) \sim Q(1,R,R^2)$, resulting in   the integrability relation in Eq.~(\ref{in1}). 
However in case of $M^2 \ll E_J^2 R'^2$, the integrated jet function is be obtained from the convolution of the standard-like jet function and the soft function~\cite{Ellis:2010rwa}, where the standard-like jet function has the same UV behavior as the standard jet function. A concrete discussion about these differences can be found in Ref.~\cite{Chien:2015cka}.

For the jet splitting contribution, at least one particle in the final state should not be included in the jet. It therefore can be written as 
\bea\label{out} 
D_{J_k/q}^{\rm{out}}(z;p_+R'/2)  &=& \int^{\infty}_{\Lambda^2} dM^2  \frac{z^{D-3}}{2N_c} \\
&&~~~\times \sum_{X_{\notin J}}\mr{Tr} \langle 0 | \delta \Bigl(\frac{p_J^+}{z}-\mc{P}_+\Bigr)\delta (M^2 - \mc{P}^{2}) \nn \Psi_n | J_k X_{\notin J}  \rangle \langle J_k X_{\notin J}| \bar{\Psi}_n | 0 \rangle_, \nnb
\eea
where $p_+ = p_J^+/z \sim 2E = 2E_J/z$ is two times of mother parton's energy. At NLO we can have at most two particles in the final state, so we can further separate this contribution as quark or gluon jet contributions. For the quark jet contribution, the gluon should be outside the jet, and vice versa for the gluon jet.  

First let us consider the quark jet contribution, where the momentum of the final state quark is given by $p_J$. 
In this case the gluon outside the jet becomes soft as $z$ goes to 1, leading to an IR singularity in the naive collinear computation unless we subtract the zero-bin contribution~\cite{Manohar:2006nz}. 
In order to isolate the singularity as $z\to 1$, we can write the quark jet contribution as follows: 
\be\label{outquark} 
D_{J_q/q}^{\rm{out}}(z;ER') = \delta (1-z) \left(\int^1_0 dz' D_{\rm{out}}^q (z';E_JR')\right) + \Bigl[D_{\rm{out}}^q (z;ER')\Bigr]_{+}.
\ee
Here the second term follows the standard plus distribution and is free of IR divergences as $z\to 1$. 

\begin{figure}[t]
\begin{center}
\includegraphics[height=5cm]{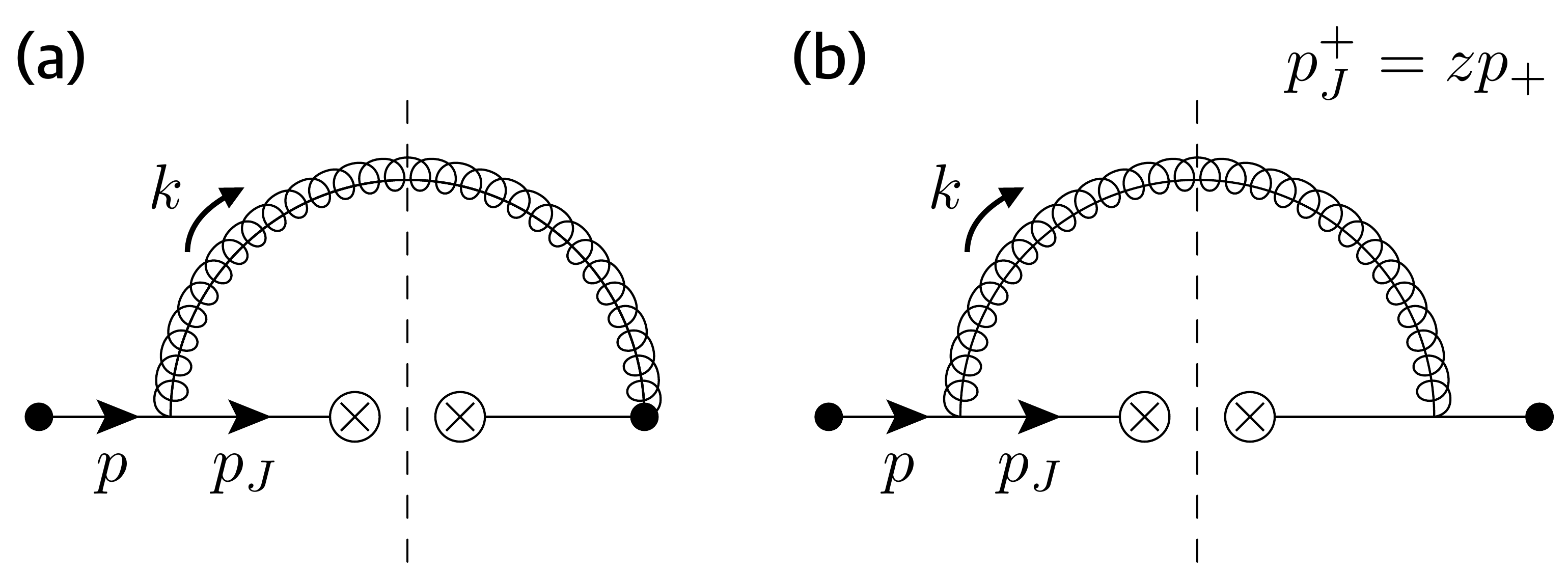}
\end{center}
\vspace{-0.3cm}
\caption{\label{fig2} \baselineskip 3.0ex 
Feynman diagrams for quark jet splitting contribution at NLO in $\as$. Here the dashed lines represent the unitary cuts. The gluon in the final state is outside the jet. Diagram (a) has its Hermitian conjugate contribution.
}
\end{figure}

In Fig.~\ref{fig2} we show the quark jet splitting contributions diagrammatically where the gluon in the final state cannot be merged into a quark jet with momentum $p_J$. The contribution of Fig.~\ref{fig2}-(a) is given by 
\bea
D^{\rm{out},(a)}_{J_q/q} &=& 4\pi g^2 C_F \mu^{2\veps}_{\overline{\mr{MS}}} \int^{\infty}_{\Lambda^2}dM^2 \frac{p_J^+}{M^2} \frac{z^{D-3}}{1-z} \int \frac{d^D k}{(2\pi)^D} \delta(k^2) \delta\left(\frac{1-z}{z} p_J^+ - k_+ \right)\delta(M^2-p_J^+ k_-), \nnb \\
\label{jsca} 
&=& \frac{\as C_F}{2\pi} \frac{(\mu^2 e^{\gamma})^{\veps}}{\Gamma(1-\veps)} \int^{\infty}_{\Lambda^2} \frac{dM^2}{(M^2)^{1+\veps}} z^{1-\veps}(1-z)^{-1-\veps}, 
\eea
where $\mu^{2}_{\overline{\mr{MS}}} = \mu^2 e^{\gamma}/(4\pi)$, and 
$\Lambda$ is the maximal jet mass for $\theta = R'$, 
\be\label{mjkT}
\Lambda^2 = p_J^{+2} t^2 \frac{1-z}{z} = p_+^2 t^2 z(1-z).
\ee
As $z$ become close to 1, Eq.~(\ref{jsca}) has IR divergence arising from soft gluon radiation. It is cancelled by the subtraction of the zero-bin contributions. 
The diagram Fig.~\ref{fig2}-(b) gives 
\bea
D^{\rm{out},(b)}_{J_q/q} &=& 4\pi g^2 C_F \mu^{2\veps}_{\overline{\mr{MS}}} (1-\veps) \int^{\infty}_{\Lambda^2}dM^2 \frac{z^{D-3}k_+}{M^2} \int \frac{d^D k}{(2\pi)^D} \delta(k^2) \delta\left(\frac{1-z}{z} p_J^+ - k_+ \right)\delta(M^2-p_J^+ k_-) \nnb \\
\label{jscb} 
&=& \frac{\as C_F}{2\pi} \frac{(\mu^2 e^{\gamma})^{\veps}}{\Gamma(1-\veps)} (1-\veps) 
\int^{\infty}_{\Lambda^2} \frac{dM^2}{(M^2)^{1+\veps}} z^{-\veps}(1-z)^{1-\veps}.
\eea
Including the Hermitian conjugate of diagram Fig.~\ref{fig2}-(a), the final result for the jet splitting is  $D_{\mr{out}}^q = 2D_{\mr{out}}^{q,(a)}+D_{\mr{out}}^{q,(b)}$. 

To calculate the part of Eq.~(\ref{outquark}) proportional to $\delta(1-z)$, we integrate over $z$, 
\bea\label{deltamout} 
\int^1_0 dz D_{\rm{out}}^q (z;E_JR') &=& -\frac{\as C_F}{2\pi} \Biggl[\frac{1}{\UV^2}+\frac{1}{\UV}\Bigl(\frac{3}{2} +\ln\frac{\mu^2}{p_{J}^{+2}t^2}\Bigr) \\ 
&&~~~~~~~~~~~
+\frac{3}{2}\ln\frac{\mu^2}{p_{J}^{+2}t^2}+\frac{1}{2}\ln^2\frac{\mu^2}{p_{J}^{+2}t^2}+\frac{13}{2}-\frac{3\pi^2}{4} \Biggr]\ .\nnb
\eea
Note that in some sense this result is trivial, since the integration of the standard jet function in Eq.~(\ref{jetf}) gives the result when there is no restriction of the phase space for the final state. 
Because Eq.~(\ref{deltamout}) is the same as the integrated jet function for the case $\theta > R'$, combining it with $\mc{J}_q (E_JR';\theta<R')$ in Eq.~(\ref{in1}) we must have 
\be\label{intjet} 
\mc{J}_q (E_JR',\theta>R')+\mc{J}_q (E_JR',\theta<R') = \int^{\infty}_0  dM^2 J_{q}(M^2) = 1. 
\ee
Thus Eq.~(\ref{deltamout}) must have the same result up to a relative minus sign compared with the first order corrections to $\mc{J}_q (E_JR';\theta<R')$, obtained from Eqs.~(\ref{in1}) and (\ref{MinkT}).

The remaining contribution  
 $[D^{\mr{out}}_{J_q/q} (z)]_+$,  is 
\bea
[D^{\mr{out}}_{J_q/q} (z)]_+ &=& [2D^{\mr{out},(a)}_{J_q/q} (z)+D^{\mr{out},(b)}_{J_q/q}(z)]_+ \nnb \\
\label{Mqjff}
&=& \frac{\as C_F}{2\pi} \Biggl[\frac{1+z^2}{1-z} \left(\frac{1}{\UV} + \ln\frac{\mu^2}{p_+^2 t^2} - 2\ln z(1-z)\right)-(1-z)\Biggr]_{+}.
\eea
Combining these results, we arrive at $D^{(1)}_{J_q/q}$, i.e., the one loop correction to the quark parton to quark jet fragmentation. Using the identity for the plus distribution,
\be\label{+iden} 
[g(z)h(z)]_+ = [g(z)]_+ h(z)  - \delta(1-z) \int^1_0 dy g(y) \Bigl[h(y)-h(1)\Bigr], 
\ee
we rewrite the renormalized NLO result  as 
\bea\label{Dqqj} 
D_{J_q/q} (z,\mu;ER') &=& \delta(1-z) + \frac{\as C_F}{2\pi} \Biggl\{ \delta(1-z) \left(\frac{3}{2} \ln \frac{\mu^2}{p_+^2 t^2}+\frac{13}{2}-\frac{2\pi^2}{3}\right)-(1-z) \\
&&+(1+z^2)\Biggl[\frac{1}{(1-z)_+}\left(\ln \frac{\mu^2}{p_+^2 t^2}-2\ln z\right)-2\left(\frac{\ln(1-z)}{1-z}\right)_+\Biggr]\Biggr\}.\nnb
\eea

\begin{figure}[t]
\begin{center}
\includegraphics[height=5cm]{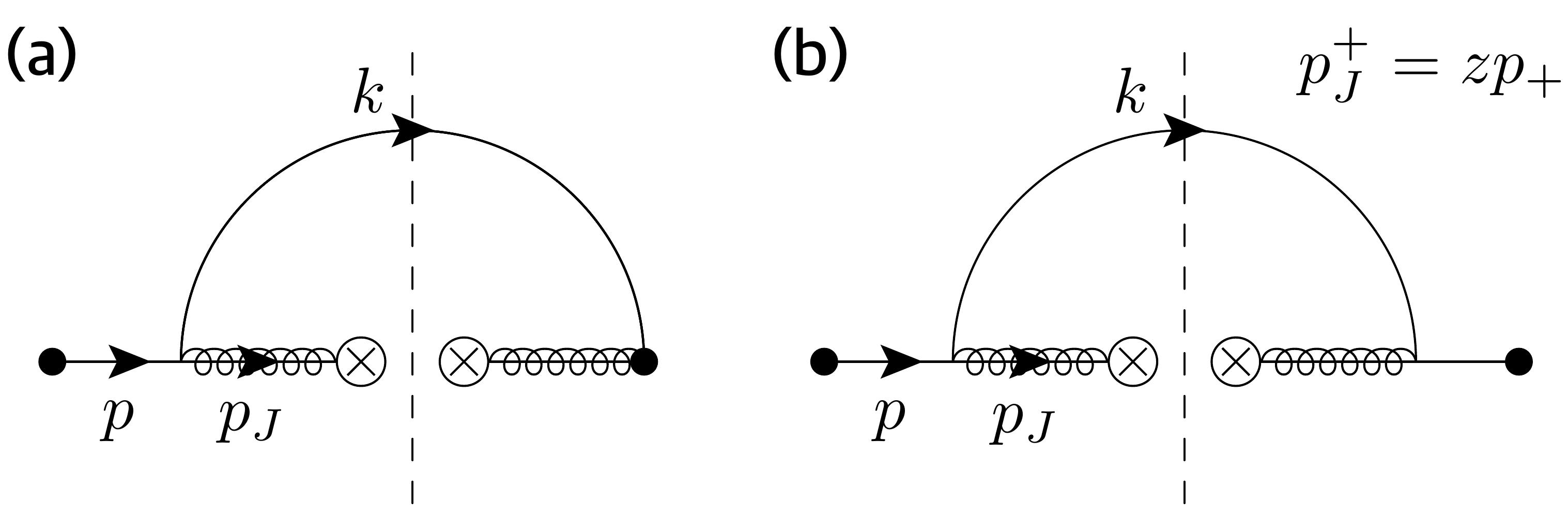}
\end{center}
\vspace{-0.3cm}
\caption{\label{fig3} \baselineskip 3.0ex 
Feynman diagrams for gluon jet splitting contribution at NLO in $\as$. Here the dashed lines represent the unitary cuts. The quark in the final state is outside the jet. Diagram (a) has its Hermitian conjugate contribution.
}
\end{figure}

We can also compute the contribution for quark parton to gluon jet fragmentation shown in Fig.~\ref{fig3}. 
In this case the gluon in the final state has the momentum $p_J^+=zp_+$ and the quark outside the jet has $(1-z)p_+$. Therefore the one loop amplitude for $z\neq 1$ should satisfy the relation $D^{\rm{out}}_{J_g/q} (z) = D^{\rm{out}}_{J_q/q} (1-z)$. 
Thus the renormalized gluon jet fragmentation function can be written down immediately,  
\be\label{Dgqj}
D_{J_g/q} (z,\mu;ER') = \frac{\as C_F}{2\pi} \Biggl[\frac{1+(1-z)^2}{z} \left(\ln\frac{\mu^2}{p_+^2 t^2} - 2\ln z(1-z)\right)-z\Biggr]\ .
\ee

The NLO result for the quark to inclusive FFJ is $D_{J/q} = D_{J_q/q} + D_{J_g/q}$, combining Eqs.~(\ref{Dqqj}) and (\ref{Dgqj}). It  satisfies the momentum sum rule shown in Eq.~(\ref{mcon}) explicitly.  
Note that here we expressed the fragmentation functions in terms of $\ln (\mu^2/p_+^2 t^2)$ rather than $\ln (\mu^2/p_J^{+2} t^2)$. If we rewrite the fragmentation functions with $\ln (\mu^2/p_J^{+2} t^2)$ using the relation $p_+=p_J^+/z$, these functions cannot satisfy the sum rule in Eq.~(\ref{mcon}) due to additional terms of $\ln z$. This fact indicates that the typical scale for the fragmentation function necessary to minimize the large logarithms with small $R$ is not $p_J^+ t\sim E_J R'$ but $p_+ t \sim E R'$.
For $z\sim \mO(1)$, the scale choice for FFJ between $E R$ and $E_J R$ might not be significant. However the proper choice of the scale can be critical in the small $z$ limit. 
 
\subsection{NLO Calculation of Gluon FFJ}
\label{subsec:gjffnlo}
As was done for the quark FFJ, we separate the NLO contributions into jet merging and jet splitting contributions. The jet merging contribution is proportional to $\delta(1-z)$ and includes the virtual contribution. Similarly to Eq.~(\ref{in1}), the jet merging contribution can be expressed as 
\be\label{ingJFF} 
D^{\rm{in}}_{J/g}(z;E_JR')  = \delta(1-z) \int^{\Lambda^2}_0  dM^2 J_{g}(M^2;\theta<R') = \delta(1-z)\mc{J}_g (E_JR';\theta<R'), 
\ee
where $\mc{J}_g$ is the integrated gluon jet function, which to NLO is given by \cite{Ellis:2010rwa,Cheung:2009sg,Chay:2015ila}
\bea\label{intJg} 
\mc{J}_g (E_JR';\theta<R')&=& 1+\frac{\as C_A}{2\pi} \Biggl[\frac{1}{\UV^2}+\frac{1}{\UV}\Bigl(\frac{\beta_0}{2C_A} +\ln\frac{\mu^2}{p_{J}^{+2}t^2}\Bigr) +\frac{\beta_0}{2C_A}\ln\frac{\mu^2}{p_{J}^{+2}t^2}\\ 
&&~~~~~~~~~~~
+\frac{1}{2}\ln^2\frac{\mu^2}{p_{J}^{+2}t^2}+\frac{67}{9}-\frac{23n_f}{18C_A}-\frac{3\pi^2}{4} \Biggr]\ ,\nnb
\eea 
where $C_A = N_c = 3$, and $\beta_0 = 11N_c/3-2n_f/3$ is the first coefficient of beta function and $n_f$ is the number of flavors. 

\begin{figure}[t]
\begin{center}
\includegraphics[height=4.5cm]{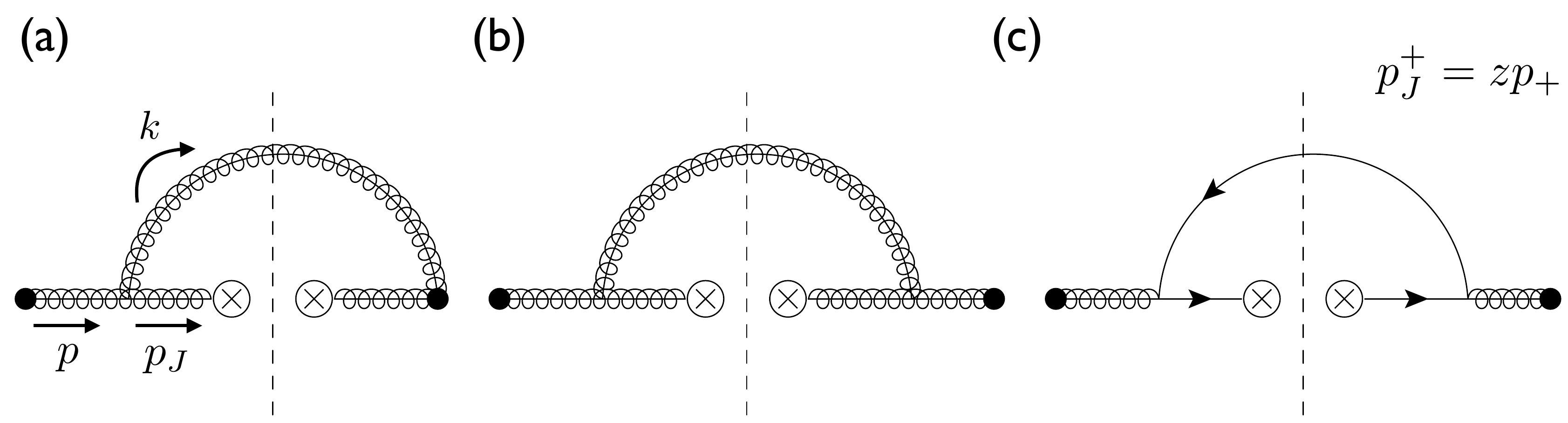}
\end{center}
\vspace{-0.3cm}
\caption{\label{gjetfrag} \baselineskip 3.0ex 
Feynman diagrams for jet splitting contributions to jet fragmentation initiated by gluon parton. Diagram (a) has its Hermitian conjugate contribution. Diagram (a) and (b) represents $g\to J_g$ splitting, and Diagram (c) represents $g\to J_q$ splitting.
}
\end{figure}

In fig.~\ref{gjetfrag}, Feynman diagrams for the jet splitting contributions are presented.\footnote{When we computed Feynman diagrams, we applied the background field method~\cite{Abbott:1980hw}.} The contribution of Diagram \ref{gjetfrag}-(a), including the zero-bin subtraction, is 
\bea\label{gluonjsca}
D^{\rm{out},(a)}_{J_g/g} &=& \frac{\as C_A}{2\pi} \Biggl\{\delta(1-z) \Bigl[-\frac{1}{2\UV^2} - \frac{1}{2\UV} \ln\frac{\mu^2}{p_J^{+2}t^2}-\frac{1}{4}  \ln^2\frac{\mu^2}{p_J^{+2}t^2} + \frac{\pi^2}{24}\Bigr]  \\
~~~~~~&&+\Bigl(\frac{1}{\UV} +\ln\frac{\mu^2}{p_+^{2}t^2}\Bigr)\Bigl[\frac{z}{(1-z)_+}+\frac{1-z}{z}+\frac{1}{2}\Bigr] \nnb \\
~~~~~~&&-2\Bigl[\frac{z\ln z}{(1-z)_+}+z\Bigl(\frac{\ln(1-z)}{1-z}\Bigr)_+ + \ln [z(1-z)]\Bigl(\frac{1-z}{z}+\frac{1}{2}\Bigr)\Bigr]\Biggr\}\ .\nnb
\eea
The contributions of Diagram \ref{gjetfrag}-(b) are given by
\be\label{gluonjscb}
D^{\rm{out},(b)}_{J_g/g} = \frac{\as C_A}{2\pi} \left(\frac{1}{\UV}+ \ln\frac{\mu^2}{p_+^{2}t^2}-2\ln[z(1-z)]\right)
\Bigl(2z(1-z)-1\Bigr)\ .
\ee
Combining Eqs.~(\ref{ingJFF}), (\ref{gluonjsca}), and (\ref{gluonjscb}), we find NLO result of gluon jet framentation function from the gluon,
\bea
D_{J_g/g}(z,\mu;ER') &=& D^{\rm{in}}_{J/g}+2D^{\rm{out},(a)}_{J_g/g}+D^{\rm{out},(b)}_{J_g/g}-\mr{UV~counter~terms}\nnb \\ 
&=& \delta(1-z) + \frac{\as C_A}{2\pi} \Biggl\{\delta(1-z)\Bigl[\frac{\beta_0}{2C_A}\ln\frac{\mu^2}{p_{J}^{+2}t^2}+\frac{67}{9}-\frac{23n_f}{18C_A}-\frac{2\pi^2}{3}\Bigr]\nnb\\
\label{gjetgfnlo}
&&~~~+2\ln\frac{\mu^2}{p_+^2t^2}\Bigl[\frac{z}{(1-z)_+}+\frac{1-z}{z}+z(1-z)\Bigr] \\
&&~~~-4\Bigl[\frac{z\ln z}{(1-z)_+}+z\Bigl(\frac{\ln(1-z)}{1-z}\Bigr)_+ + \ln [z(1-z)]\Bigl(\frac{1-z}{z}+z(1-z)\Bigr)\Bigr]\Biggr\}\ .\nnb
\eea 

Diagram \ref{gjetfrag}-(c) contributes to the quark jet fragmentation. The one loop result is given by
\bea
D_{J_q/g}(z,\mu;ER') &=&  D^{\rm{out},(c)}_{J_q/g}-\mr{UV~counter~terms} \nnb \\ 
\label{gjetqfnlo}
&=& \frac{\as}{2\pi} \Bigl[\Bigl(\ln\frac{\mu^2}{p_+^2t^2}-2\ln [z(1-z)]\Bigr)\frac{z^2+(1-z)^2}{2}-z(1-z)\Bigr]\ .
\eea 
Note that that Eqs.~(\ref{gjetgfnlo}) and (\ref{gjetqfnlo}) satisfy the momentum conservation sum rule in Eq.~(\ref{mcon}),
\be\label{gsumr} 
\int_0^1 dz z \Big([D_{J_g/g} (z) + n_f D_{J_q/g} (z) + n_f  D_{J_{\ov{q}}/g} (z)\Bigr)
=\int_0^1 dz z \Big([D_{J_g/g} (z) + 2n_f D_{J_q/g} (z) \Bigr) = 1. 
\ee

\subsection{Renormalization Scaling Behavior}
As can be seen in Eqs.~(\ref{Dqqj}), (\ref{Dgqj}), (\ref{gjetgfnlo}), and (\ref{gjetqfnlo}), the renormalization group (RG) scaling behavior of the FFJs follows the well-known DGLAP evolution,
\be\label{DGLAP} 
\frac{d}{d\ln\mu} D_{J_l/k} (x,\mu) = \frac{\as(\mu)}{\pi}\int_x^1 \frac{dz}{z} P_{lm}(z) D_{J_m/k} (x/z,\mu),  
\ee
where the leading splitting kernels are given by 
\bea 
\label{pqq}
P_{qq}(z) &=& C_F \Bigl[\frac{3}{2}\delta(1-z) + \frac{1+z^2}{(1-z)_+} \Bigr], \\
\label{pgq}
P_{gq}(z) &=& C_F \Bigl[\frac{1+(1-z)^2}{z}\Bigr], \\
\label{pqg}
P_{qg}(z) &=& \frac{1}{2} \Bigl[z^2+(1-z)^2], \\
\label{pgg}
P_{gg}(z) &=& \frac{\beta_0}{2}\delta(1-z)+2C_A\Bigl[\frac{z}{(1-z)_+}+\frac{1-z}{z}+z(1-z)\Bigr]\ . 
\eea

When we compare the higher order result of the FFJ with the fragmentation of a massless parton, the size of the jet, $ER'$, suppresses IR sensitivity of the FFJ while the latter has IR divergences. However, both have identical UV behaviors, since the UV divergences arise when the splitting of two particles becomes hard with given large splitting angle.

Comparing to other work, we find that our NLO results for FFJ in Eqs.~(\ref{Dqqj}), (\ref{Dgqj}), (\ref{gjetgfnlo}), and (\ref{gjetqfnlo}) are the same as ``jet functions", $j_{k\to l}$, in Ref.~\cite{Kaufmann:2015hma}, where the only difference is that the logarithmic terms has been expressed not as $ER'$ but $E_JR'= ER'/z$. This removes the $\ln z$ term in our expression. However, if we write it this way, we cannot guarantee the momentum sum rule in Eq.~(\ref{mcon}) as we mentioned before. That might give some subtleties for the comparison with other approaches to the estimation of FFJ at  higher orders~\cite{Dasgupta:2014yra,Dasgupta:2016bnd}.  

As noted in the introduction, while completing this work, Ref~\cite{Kang:2016mcy} appeared on the arXiv. The authors have also computed the FFJ at NLO using SCET. The results are the same as ours, but they have the same expression as appearing in Ref.~\cite{Kaufmann:2015hma}. They claimed that all the virtual diagrams vanish because they are scaleless. However, we believe it is important to carefully separate the UV and IR divergences to obtain a clear picture of the physics. For example, for the case of the jet merging (in-jet) contribution, only when we combine the virtual and real contributions can we obtain an IR finite result.

\section{Factorization Theorem for the Fragmentation inside a Jet}
\label{Facincjet}

To begin, let us consider the scattering cross section with a HFF at a hadron collider:
\bea 
\sigma &=& \sum_k \int dwdydp_T~\frac{d\sigma_k}{dydp_T}~D_{H/k} (w) \nnb \\
&=& \sum_k \int dw dy dp_T  dp_T^H~\frac{d\sigma_k}{dydp_T}~ \delta(w p_T^k -p_T^H)D_{H/k} (w), 
\eea
where $\sigma_k$ is the scattering cross section for the inclusive process with a final parton $k$, $N_1N_2 \to k X$, $p_T$ is the transverse momentum of the parton $k$ to beam axis, $y$ is the rapidity of the parton $k$, and the rapidity of the hadron can be approximated to be the same as the parton. The differential scattering cross section for the hadron $H$ is
\be\label{xsec1} 
\frac{d\sigma}{dydp_T^H} = \sum_k \int^1_{x_H} \frac{dw}{w} \frac{d\sigma_k(y,x_H/w)}{dydp_T} D_{H/k} (w),
\ee
where $x_H = p_T^H/Q_T$ and so $x_H/w = p_T/Q_T$, with $Q_T$ being the  maximal possible $p_T$ at a given rapidity.

Next we would like to consider the fragmentation of the hadron inside a jet. In order to do this we factorize the inclusive HFF, 
\be\label{FFfac} 
D_{H/k} (w) = \sum_l \int^1_w \frac{dz}{z} B_{J_l/k}\left(\frac{w}{z};ER'\right) \tilde{D}_{H/J_l} (z;E_J R'),
\ee 
where $J_l$ is the jet with a parton $l$, and the momentum fractions are defined as $z=p_H^+/p_J^+=p_T^H/p_T^J$ and $p_J^+/p_+=p_T^J/p_T=w/z$. $B_{J_l/k}$ is the jet splitting kernel from the parton $k$, and 
$\tilde{D}_{H/J_l}$ is the hadron fragmentation from $J_l$. 

$\tilde{D}_{H/J_l}$ can be computed by the integration of the fragmenting jet function (FJF)~\cite{Procura:2009vm,Jain:2011xz},
\be\label{eFF} 
\tilde{D}_{H/J_l}(z;E_J R') = \int^{\Lambda^2}_0 dM^2 J_{H/l} (z,M^2),
\ee
where the LO parton level FJF is normalized as $J_{m/l}(z,M^2) = \delta(1-z)\delta(M^2) \delta_{ml}$.  
$\Lambda^2$ is the maximum  jet mass with a given hadron energy fraction $z$. For a $\mr{k_T}$-type jet algorithm, it can be expressed as 
\be
\label{maxjmass} 
\Lambda_{\mr{k_T}}^2 = z(1-z) p_J^{+2} \tan^2\Bigl(\frac{R'}{2}\Bigr). 
\ee
The computation of $\tilde{D}_{H/J_l}(z)$ at NLO was  done in Ref.~\cite{Procura:2011aq,Chien:2015ctp}. We also show the NLO calculation in appendix~\ref{A1} separating the UV and IR divergences carefully.

\begin{figure}[t]
\begin{center}
\includegraphics[height=10cm]{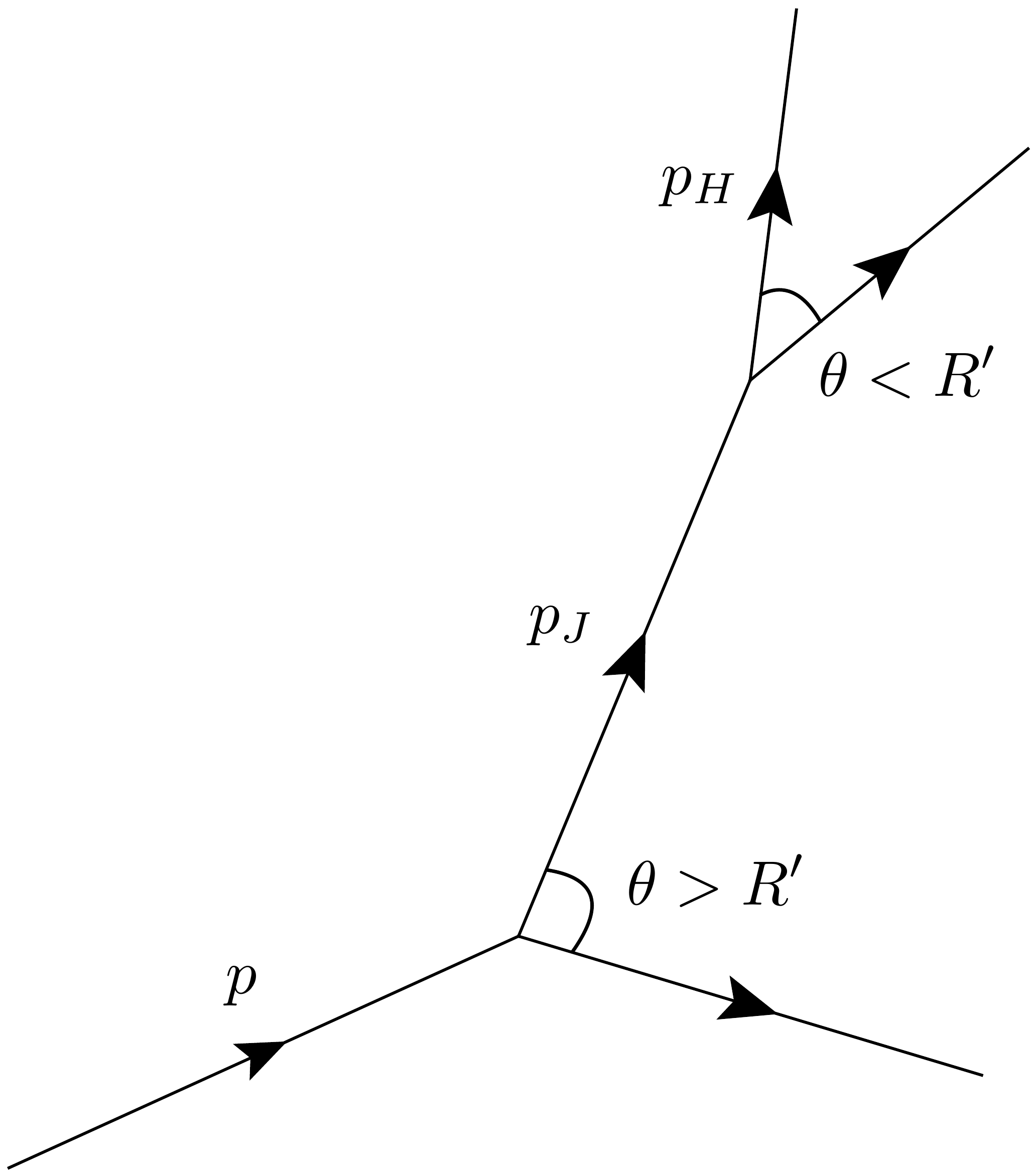}
\end{center}
\vspace{-0.3cm}
\caption{\label{fig4} \baselineskip 3.0ex 
Fragmentation process from the parton $(p)$ to the hadron $(p_H)$ through the jet $(p_J)$. }
\end{figure}

$B_{J_l/k}$ is the jet splitting kernel from the mother parton $k$. If we consider the process $k\to lm$, the contribution to $B_{J_l/k}$ comes from the case where the angle between the partons $l$ and $m$ is larger than $R'$. Because the convolution of $B_{J_l/k}$ and $\tilde{D}_{H/J_l}$ includes all  possible fragmentation processes, the result should be the same as the inclusive HFF. However Eq.~(\ref{FFfac}) shows that it is possible to describe the whole fragmentation process with a more exclusive observable. The perturbative result of $B_{J_l/k}$ can be obtained from the matching between $D_{H/k}$ and $\tilde{D}_{H/J_l}$. 
In Fig.~\ref{fig4} we show the fragmentation process of the hadron through a jet schematically.

We can also consider the direct computation of $B_{J_l/k}$ based on the calculation of the FFJ in the previous section. 
From the description above, $B_{J_l/k}$ should be 
\be\label{Bjlk} 
B_{J_l/k} (z,\mu;ER') = \delta(1-z)\delta_{lk} +D^{\rm{out}}_{J_l/k}(z,\mu;ER'),
\ee
where $D^{\rm{out}}_{J_l/k}(z)$ is the jet splitting (out-jet) contribution considered in the FFJ calculation. The typical scale for the jet splitting is  $p_+ \tan(R'/2)$, which can be approximated as $ER'$. Interestingly we find that the perturbative result of the FFJ can be obtained if the higher order result for the jet merging (in-jet) contribution is added to Eq.~(\ref{Bjlk}). As shown in Eqs.~(\ref{in1}) and (\ref{ingJFF}), the jet merging contribution can be  expressed as $D^{\rm{in}}_{J/k}=\delta(1-z) \mc{J}_k(E_J R')$. Therefore perturbatively we have the relation~\footnote{This result has been used for the factorization of the jet mass distribution~\cite{Idilbi:2016hoa}.} 
\be\label{JFFB} 
D_{J_l/k} (z,\mu; ER') =  B_{J_l/k} (z,\mu;ER') \mc{J}_l(\mu;E_JR'), 
\ee 
where the index of the jet parton, $l$, is not summed over. Note that the factorized result in Eq.~(\ref{JFFB}) has been only confirmed to one loop order. To validate this result beyond  NLO, we would need to  check the two loop calculation explicitly, which is beyond the scope of this paper. 

On the right side of Eq.~(\ref{JFFB}), having  $\mc{J}_l$ rather than $\mc{J}_k$  makes sense beyond NLO accuracy. To see this, consider the case with three final partons at NNLO. If all three particles combine into the jet, the contribution to the FFJ is proportional to $\delta(1-z)$. As seen in Eq.~(\ref{Bjlk}), the $\delta_{lk}$ in $B_{J_l/k}$ guarantees the jet merging contribution is the integrated jet function for the parton $k$. However, if we consider the process $k \to lm\to (l_1l_2)m$ where $l\to l_1l_2$ merged in the jet, this NNLO contribution can be expressed as the multiplication of $M_{J_l/k}^{\mr{out},(1)}$ and $\mc{J}^{(1)}_l$.\footnote{\baselineskip 3.0ex
We have not considered three parton splitting processes at NNLO explicitly. (For the details, see Ref.~\cite{Catani:1998nv}.)  
It may complicate the factorization in Eq.~(\ref{JFFB}). } 
 Here the superscript (1) denotes the contributions at NLO.  

If we apply the momentum conservation sum rule for the hadron to $\tilde{D}_{H/J_l}(z)$ in Eq.~(\ref{eFF}), we obtain~\cite{Procura:2011aq}
\be\label{sumD}
\sum_H \int^1_0 dz z \tilde{D}_{H/J_l} (z) = 
\int^{\Lambda^2}_0 dM^2 \sum_H \int^1_0 dz z J_{H/l} (z,M^2) = \int^{\Lambda^2}_0 dM^2 J_l(M^2) = \mc{J}_l.
\ee 
This also implies the relation of Eq.~(\ref{JFFB}). As denoted in Eq.~(\ref{mcon}),  the FFJ satisfies the sum rule. Therefore, when applied to Eq.~(\ref{FFfac}), the sum rule for the inclusive HFF is guaranteed,   
\bea 
\sum_H \int^1_0 dw w D_{H/k} (w) &=& \sum_l \int^1_0 dx x B_{J_l/k}(x) \sum_H \int^1_0 dz z \tilde{D}_{H/J_l} (z)
\nnb \\
&=& \sum_l \int^1_0 dx x B_{J_l/k}(x) \mc{J}_l =\sum_l \int^1_0 dx x D_{J_l/k} = 1.
\eea 

From Eq.(\ref{sumD}), we see that the normalization of $\tilde{D}_{H/J_l}$ is not adequate for a probability. Dividing $\tilde{D}_{H/J_l}$ by the integrated jet function, we can introduce the HFF inside a jet~\cite{Chien:2015ctp}\footnote{In Ref.~\cite{Chien:2015ctp}, this HFF inside the jet has been called as a jet fragmentation function. } 
\be
\label{HFFinJ}
D_{H/J_l} (z;ER') = \frac{\tilde{D}_{H/J_l} (z,\mu;ER')}{\mc{J}_l(\mu;E_JR')}\ .
\ee   
Note that this HFF inside the jet has no renormalization scale dependence because the scale dependence for $\tilde{D}_{H/J_l}$ is cancelled by $\mc{J}_l$.  (This can be seen by considering the scale dependence in Eq.~(\ref{dxsec3}) below.)
Finally combining Eqs.~(\ref{JFFB}) and (\ref{HFFinJ}) we can rewrite Eq.~(\ref{FFfac})  as 
\be\label{HFFfact} 
D_{H/k} (w,\mu) = \sum_l \int^1_w \frac{dz}{z} D_{J_l/k}\left(\frac{w}{z},\mu;ER'\right) D_{H/J_l} (z;E_JR').
\ee

Like a hadron, a jet is also an observable. So it is useful to consider the differential scattering cross section observing the jet and hadron simultaneously. To derive the factorization theorem we combine Eq.~(\ref{xsec1}) with Eq.~(\ref{HFFfact})   
\bea
\frac{d\sigma}{dydp_T^H} &=& \sum_{k,l} \int^1_{x_H} \frac{dw}{w} \frac{d\sigma_k(y,x_H/w)}{dydp_T} \nnb \\
\label{xsec2}
&&\times 
\int^1_w \frac{dz}{z} \int dp_T^J \delta (x_J Q_T - p_T^J) 
D_{J_l/k}\left(\frac{w}{z}\right) D_{H/J_l} (z),
\eea
where $x_J = p_T^J/Q_T$, and we put in the identity $1=\int dp_T^J \delta (x_J Q_T - p_T^J)$. The delta function becomes
\be\label{delta1} 
\delta(x_J Q_T- p_T^J) =  \frac{1}{Q_T}\delta \left(x_J-\frac{x_H}{z}\right)  
= \frac{z^2}{x_H Q_T} \delta\left(z-\frac{x_H}{x_J}\right)\ .
\ee
Therefore the differential scattering cross section for the jet and the hadron inside the jet can be written as 
\bea
\frac{d\sigma}{dydp_T^J dp_T^H} &=& \sum_{k,l}\int^1_{x_H} \frac{dw}{w} \frac{d\sigma_k(y,p_T/Q_T)}{dydp_T}
\frac{z}{x_H Q_T} D_{J_l/k}\left(\frac{p_T^J}{p_T}\right) D_{H/J_l} (z)\nnb \\
\label{dxsec1}
&=& \sum_{k,l}\int^1_{x_J} \frac{dx}{x} \frac{d\sigma_k(y,p_T/Q_T=x_J/x)}{dydp_T}
\frac{z}{x_H Q_T} D_{J_l/k}(x) D_{H/J_l} (z).
\eea
In the second equality we introduced the variable $x=p_T^J/p_T = w/z$, hence 
\be\label{intid} 
\int^1_{x_H} \frac{dw}{w} = \int^1_{x_J} \frac{dx}{x}\ . 
\ee
Finally we have\footnote{\baselineskip 3.0ex
In Ref.~\cite{Kaufmann:2015hma}, the similar factorization theorem has been analyzed from the full NLO calculation. We can clearly see the similarity if we express $D_{H/J_l}$ as Eq.~(\ref{facD2}) when $\mu \ll E_JR'$.}
\be\label{dxsec2}
\frac{d\sigma}{dydp_T^Jdz}= \sum_{k,l}\int^1_{x_J} \frac{dx}{x} \frac{d\sigma_k(y,x_J/x)}{dydp_T}
D_{J_l/k}(x) D_{H/J_l} (z).
\ee

The factorization theorem in Eq.~(\ref{dxsec2}) is very useful. For example, instead of the observed hadron, we can consider a subjet inside a fat jet. In this case the factorization theorem becomes 
\be\label{dxsec3}  
\frac{d\sigma}{dydp_T^Jdz}= \sum_{k,l}\int^1_{x_J} \frac{dx}{x} \frac{d\sigma_k(y,x_J/x)}{dydp_T}
D_{J_l/k}(x) D_{j/J_l} (z), 
\ee
where $z$ is the momentum fraction of the subjet $j$ compared to the fat jet $J$ given by $z=p_{j}^+/p_J^+=p_T^j/p_T^J$ and $D_{j/J_l}$ is the subjet fragramentation function inside the fat jet. We investigate this more in the following section.

\section{Subjet Fragmentation inside a Fat jet}  
\label{subjet} 

For the  description of the subjet fragmentation function (sJFF) inside a jet, $D_{j/J_l}$ in Eq.~(\ref{dxsec3}), the parton splitting within a fat jet $(J)$ with the radius $R$ only is  taken into account. It has a restricted phase space for collinear particle radiations compared to the fully inclusive FFJ. As with the HFF inside a jet defined in Eq.~(\ref{HFFinJ}), sFFJ can be written as 
\be
\label{sJFFinJ}
D_{j/J_l} (z;R'/r') = \frac{\tilde{D}_{j/J_l} (z,\mu;E_JR',R'/r')}{\mc{J}_l(\mu;E_JR')} \ , 
\ee   
where $r'$ is the maximal subjet radius. As we will see, the normalized sJFF, $D_{j/J_l}$, has no scale dependence except the coupling constant, but depends on the logarithm of $R'/r'$.

The naive unnormalized sJFF, $\tilde{D}_{j/J_l}$, is described by
\bea\label{sJFFq} 
\tilde{D}_{j_k/J_q}(z,\mu) &=& \frac{z^{D-3}}{2N_c} \sum_{X_{j-1},X_{\notin j}} \mr{Tr} \langle 0 | \delta \bigl(\frac{p_j^+}{z}-\mc{P}_+\bigr)\nn \Psi_n | j_k(p_{j}^+,r)X_{\notin j} \in J(p_J^+,R)\rangle \\
&&\times \langle j_k(p_{j}^+,r)X_{\notin j} \in J(p_J^+,R) | \bar{\Psi}_n |0\rangle, \nnb
\eea
where $j_k$ represents the subjet with parton $k$, and $r$ is its radius, $X_{j-1}$ is possible final states within the subjet except the parton $k$, and  
$X_{\notin j}$ are the final states not to be included in the subjet, but contained in the jet $J$. The gluon-initiated sJFF can be expressed similarly in terms of $\mc{B}_{n}^{\pp a}$ in the adjoint representation. 

When we consider the one loop corrections, we will separate the corrections into in-subjet and out-subjet contributions as in Sec.~\ref{incjff}. With the same reasoning as  Eq.~(\ref{MinkT}), we obtain the in-subjet contribution including the virtual corrections, 
\bea\label{sjin}
D_{\rm{in}}(z;E_Jr') &=& \delta(1-z)\Biggl\{1+\frac{\as C_F}{2\pi} \Biggl[\frac{1}{\UV^2}+\frac{1}{\UV}\Bigl(\frac{3}{2} +\ln\frac{\mu^2}{p_{J}^{+2}t_r^2}\Bigr) \\ 
&&~~~~~~~~~~~
+\frac{3}{2}\ln\frac{\mu^2}{p_{J}^{+2}t_r^2}+\frac{1}{2}\ln^2\frac{\mu^2}{p_{J}^{+2}t_r^2}+\frac{13}{2}-\frac{3\pi^2}{4} \Biggr]\Biggr\}\ ,\nnb
\eea 
where $t_r\equiv\tan(r'/2)\sim r'/2$.

\begin{figure}[t]
\begin{center}
\includegraphics[height=5cm]{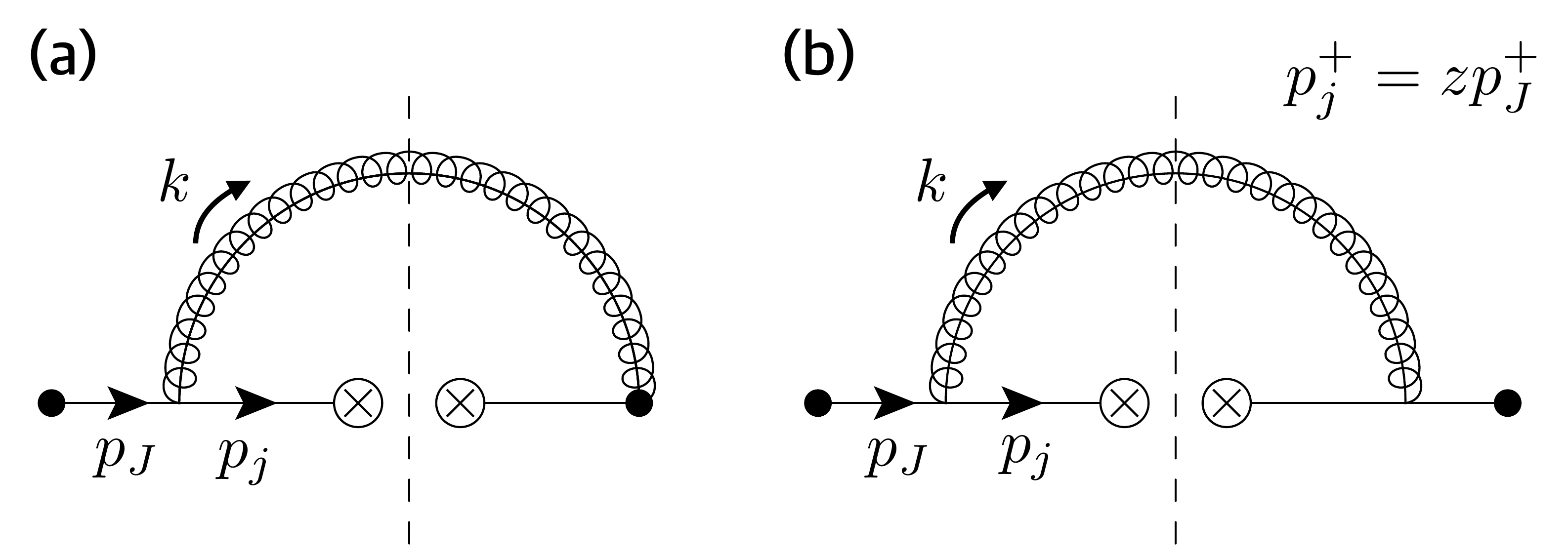}
\end{center}
\vspace{-0.3cm}
\caption{\label{fig6} \baselineskip 3.0ex 
Feynman diagrams of real gluon emissions for the subjet quark fragmentation inside a jet at NLO. Diagram (a) has its Hermitian conjugate contribution.
}
\end{figure}

The out-subjet contribution comes from real radiations with $r'<\theta<R'$. The naive collinear contribution from the Feynman diagram in Fig.~\ref{fig6}-(a) is
\bea\label{sjqa} 
\tilde{D}_{\mr{out}}^{(a)}(z) &=& \frac{\as C_F}{2\pi} \frac{(\mu^2 e^{\gamma})^{\veps}}{\Gamma(1-\veps)} \int^{p_J^{+2}t_R^2 z(1-z)}_{p_J^{+2}t_r^2 z(1-z)} \frac{dM^2}{(M^2)^{1+\veps}} z^{1-\veps}(1-z)^{-1-\veps}\\
&=& \tilde{I}_{\mr{out}}^{(a)} \delta(1-z) + \Bigl[D_{\mr{out}}^{(a)}(z)\Bigr]_{+}\ , \nnb
\eea 
where $t_R\equiv\tan(R'/2)\sim R'/2$, and the tilde represents the result before zero-bin subtractions. $\tilde{I}_{\mr{out}}^{(a)}$ is can be extracted by integrating over $z$,  
\be 
\tilde{I}_{\mr{out}}^{(a)} = \int^1_0 dz \tilde{D}_{\mr{out}}^{(a)}(z) 
\label{sjIqa}
= \frac{\as C_F}{2\pi} \Biggl[\Bigl(\frac{1}{2\IR} +1\Bigr)\ln\frac{t_r^2}{t_R^2}+\frac{1}{4}\Bigl(\ln^2\frac{\mu^2}{p_J^{+2}t_R^2}-\ln^2\frac{\mu^2}{p_J^{+2}t_r^2}\Bigr)\Biggr]\ . 
\ee
Here IR divergence arises as $z \to 1$, which is cancelled by the zero-bin contribution, 
\bea 
D_{\mr{out},0}^{(a)}(z) &=& \frac{\as C_F}{2\pi} \frac{(\mu^2 e^{\gamma})^{\veps}}{\Gamma(1-\veps)} \delta(1-z) \int^{\infty}_0  dk_+ k_+^{-1-\veps} \int^{t_R^2k_+}_{t_r^2 k_+} k_-^{-1-\veps} \nnb \\
\label{zbsjq}
&=& \frac{\as C_F}{2\pi} \Biggl[\frac{1}{2}\Bigl(\frac{1}{\UV}-\frac{1}{\IR}\Bigr) \ln \frac{t_R^2}{t_r^2} \Biggr]\delta(1-z).
\eea 
Hence the IR divergence in Eq.~(\ref{sjIqa}) is converted to a UV divergence by the zero-bin subtraction. $[D_{\mr{out}}^{(a)}(z)]_{+}$ is free from IR divergence as $z\to 1$ and is given by 
\be\label{Madsj}
\Bigl[D_{\mr{out}}^{(a)}(z)\Bigr]_{+}=\frac{\as C_F}{2\pi} \Bigl[\frac{z}{1-z}\Bigr]_+\ln \frac{t_R^2}{t_r^2}\ .
\ee

The out-subjet contribution from  diagram  Fig.~\ref{fig6}-(b) is 
\bea\label{sjqb} 
D_{\mr{out}}^{(b)}(z) &=& \frac{\as C_F}{2\pi} \frac{(\mu^2 e^{\gamma})^{\veps}}{\Gamma(1-\veps)} (1-\veps)\int^{p_J^{+2}t_R^2 z(1-z)}_{p_J^{+2}t_r^2 z(1-z)} \frac{dM^2}{(M^2)^{1+\veps}} z^{-\veps}(1-z)^{1-\veps}\\
&=& I_{\mr{out}}^{(b)} \delta(1-z) + \Bigl[D_{\mr{out}}^{(b)}(z)\Bigr]_{+}\ , \nnb
\eea 
where the terms in the second line are
\bea
\label{sjIqb}
I_{\mr{out}}^{(b)}&=&\int^1_0 dz D_{\mr{out}}^{(b)}(z) 
= \frac{\as C_F}{2\pi} \left(\frac{1}{2}\ln\frac{t_R^2}{t_r^2}\right), \\
\label{Mbdsj}
\Bigl[D_{\mr{out}}^{(b)}(z)\Bigr]_{+}&=&\frac{\as C_F}{2\pi} (1-z)_+\ln \frac{t_R^2}{t_r^2}\ .
\eea

Finally combining Eqs.~(\ref{sjin}), (\ref{sjIqa}), (\ref{zbsjq}), (\ref{Madsj}), (\ref{sjIqb}), and (\ref{Mbdsj}), we obtain bare NLO result for the naive sJFF:
\bea 
\tilde{D}_{j_q/J_q}(z,\mu) &=& D_{\rm{in}}(z)+2\Bigl[\tilde{D}_{\rm{out}}^{(a)}(z)-D_{\rm{out},0}^{(a)}(z)\Bigr]+D_{\rm{out}}^{(b)}(z) \nnb\\
\label{nsJFFNLO}
&=&\delta (1-z) \Biggl\{1+\frac{\as C_F}{2\pi}\Biggl[\frac{1}{\UV^2}+\frac{1}{\UV}\Bigl(\frac{3}{2}+\ln\frac{\mu^2}{p_{J}^{+2}t_R^2}\Bigr)+\frac{3}{2}\ln\frac{\mu^2}{p_{J}^{+2}t_R^2} 
+\frac{1}{2}\ln^2\frac{\mu^2}{p_{J}^{+2}t_R^2}\nnb\\ 
&&~~~~~~~~~+\frac{13}{2}-\frac{3\pi^2}{4} \Biggr]\Biggr\}
+\frac{\as C_F}{2\pi}\Bigl[\frac{1+z^2}{1-z}\Bigr]_+\ln \frac{t_R^2}{t_r^2}\  .
\eea
Therefore normalized sJFF can be written as 
\bea
D_{j_q/J_q}(z) &=& \frac{\tilde{D}_{j_q/J_q}(z;E_JR',r'/R')}{\mc{J}_q(\mu;E_JR')} =\delta(1-z)+\frac{\as C_F}{2\pi}\Bigl[\frac{3}{2}\delta(1-z)+\frac{1+z^2}{(1-z)_+} \Bigr] \ln \frac{t_R^2}{t_r^2}\nnb \\
\label{sJFFNLO} 
&=& \delta(1-z)+\frac{\as}{2\pi} P_{qq}(z) \ln \frac{t_R^2}{t_r^2}\  .
\eea

The gluon subjet framentation function from a quark jet can be easily computed. From Eqs.~(\ref{Madsj}) and (\ref{Mbdsj}), exchanging $z \leftrightarrow 1-z$ and removing `+'-distribution we obtain 
\be
D_{j_g/J_q}(z,\mu) = 2D_{\mr{out},g/q}^{(a)}(z)+D_{\mr{out},g/q}^{(b)}(z) = \frac{\as C_F}{2\pi}\frac{1+(1-z)^2}{z}\ln \frac{t_R^2}{t_r^2} =\frac{\as}{2\pi} P_{gq}(z) \ln \frac{t_R^2}{t_r^2}\ .
\ee

In a similar manner can compute the sJFFs from the gluon jet. They are given by 
\bea
\label{sJFFgg} 
D_{j_g/J_g}(z) &=& \delta(1-z)+\frac{\as}{2\pi} P_{gg}(z) \ln \frac{t_R^2}{t_r^2}\ , \\
\label{sJFFqg} 
D_{j_q/J_g}(z) &=& \frac{\as}{2\pi} P_{qg}(z) \ln \frac{t_R^2}{t_r^2}\ . 
\eea

If $t_R \gg t_r$, the perturbative series expansion fails, and we need to resum the large logarithms of $t_R/t_r$ to all order in $\as$. To do this, first we integrate out the mode with fluctuations of order $p^2\sim p_J^{+2}t_R^2$. Then, at the lower scale $\mu \sim p_J^+ t_r$, we consider the sJFF setting  the upper limit $p_J^+t_R \to \infty$. Therefore, similar to Eq.~(\ref{facD2}), we obtain the factorization theorem for the subjet fragmentation function 
\be\label{facD3} 
D_{j_l/J_k}(z;R'/r') = \int^1_z \frac{dx}{x} K_{m/k} (z/x,\mu;E_JR') D_{j_l/m} (x,\mu;E_Jr'). 
\ee 
Here $D_{j_l/m}$ is the standard FFJ for the subjet within the radius $r$ and the momentum of the mother parton is given by $p_J$. The perturbative result is the same as the result in sec.~\ref{incjff} with the replacement $E\to E_J$ and $R'\to r'$. 

The perturbative kernels $K_{m/k}$ are the matching coefficients between $D_{j_l/J_k}$ and $D_{j_l/m}$ and are the result of integrating out the short distance interactions with  offshellness  $E_J^2R'^2$. They are  
\bea
K_{q/q}(z,\mu) &=& \delta(1-z) -\frac{\as}{2\pi}\Biggl\{P_{qq} (z) \ln \frac{\mu^2}{p_J^{+2}t_R^2} 
+ C_F \Biggl[\delta(1-z) \Bigl(\frac{13}{2}-\frac{2\pi^2}{3}\Bigr)-(1-z) \nnb \\
&&-2(1+z^2)\left(\frac{\ln z}{(1-z)_+}+\Bigl(\frac{\ln(1-z)}{1-z}\Bigr)_+\right)\Biggr]\Biggr\}\ ,\\
K_{g/q}(z,\mu) &=& -\frac{\as}{2\pi} \Biggl[P_{gq}(z) \left(\ln\frac{\mu^2}{p_J^{+2}t_R^2}  - 2\ln z(1-z)\right)-z C_F\Biggr]\ ,\\
K_{g/g}(z,\mu) &=& \delta(1-z) -\frac{\as}{2\pi}\Biggl\{P_{gg} (z) \ln \frac{\mu^2}{p_J^{+2}t_R^2} +C_A\Biggl[\delta(1-z)\Bigl(\frac{67}{9}-\frac{23n_f}{18C_A}-\frac{2\pi^2}{3}\Bigr)\nnb\\
&&-4\Bigl[\frac{z\ln z}{(1-z)_+}+z\Bigl(\frac{\ln(1-z)}{1-z}\Bigr)_+ + \ln [z(1-z)]\Bigl(\frac{1-z}{z}+z(1-z)\Bigr)\Bigr]\Biggr]\Biggr\}\ ,\\
\label{Kqg}
K_{q/g}(z,\mu) &=& -\frac{\as}{2\pi} \Bigl[P_{qg}(z) \Bigl(\ln\frac{\mu^2}{p_J^{+2}t_R^2}-2\ln [z(1-z)]\Bigr)-z(1-z)\Bigr]\ .
\eea

The above results are very interesting. If we replace $p_J^+$ with the mother parton's momentum, $p_+$,  we see that the NLO results of $K_{m/k}$ are the same as NLO corrections to the FFJ with a relative minus sign given in as can be seen from  from Eqs.~(\ref{Dqqj}), (\ref{Dgqj}), (\ref{gjetgfnlo}), and (\ref{gjetqfnlo}). Also, we can see that the sJFF is free from the specific momentum of mother parton,  only depending upon the momentum ratio. So, even though there is not much physical meaning, at the computation level we may rewrite Eq.~(\ref{facD3})  as $D_{j/J}(R'/r') = K(ER')\otimes D_j(Er')$, with $\otimes$ is the convolution of the momentum fraction and we show the compatible scale for each function where the compatible scale $X$ appears in $\ln(\mu^2/X^2)$ in the NLO calculation.  

Based on the results for the factorization theorem in sec.~\ref{Facincjet}, let us consider an inclusive scattering cross section for the jet, $j$ with the radius $r$ in $e^+e^-$ annihilation. The scattering cross section is schematically given by 
\be
\label{dsdE}
\left(\frac{d\sigma}{dE_j}\right)_m = \left(\frac{d\sigma}{dE}\right)_k \otimes [D_j (Er)]_{km} =  \left(\frac{d\sigma}{dE}\right)_k \otimes [D_J (ER)]_{kl} \otimes [D_{j/J} (R/r)]_{lm},
\ee
where the subscripts $k,~l$, and $m$ denote parton flavors, which are summed for the same indices. $[D_J]_{kl}$ represents $D_{J_l/k}$, and $[D_{j/J}]_{lm} = D_{j_m/J_l}$. 
As discussed below Eq.~(\ref{Kqg}) $[D_{j/J}(R/r)]_{km}= [K(E_JR)]_{kl}\otimes [D_j(E_Jr)]_{lm} = [K(ER)]_{kl}\otimes [D_j(Er)]_{lm}$, where $[K]_{kl} = K_{l/k}$. Hence Eq.~(\ref{dsdE}) can be written as  
\bea
\left(\frac{d\sigma}{dE}\right)_k \otimes [D_j (Er)]_{kn} &=& \left(\frac{d\sigma}{dE}\right)_k \otimes [D_J (ER)]_{kl} \otimes [K(ER)]_{lm}\otimes [D_j(Er)]_{mn} \nnb \\
\label{dsdE1}
&=& \left(\frac{d\sigma}{dE}\right)_k \otimes [D_J (ER)]_{kl} \otimes [D_J^{-1} (ER)]_{lm} \otimes [D_j(Er)]_{mn} \\
&=&\left(\frac{d\sigma}{dE}\right)_{m} \otimes [D_j (Er)]_{mn}. \nnb
\eea
This result implies that $K(ER)$ represents the inverse process of  jet fragmentation. This fact demonstrates our observation that the NLO correction to $K$ putting $p_+$ instead of $p_J^+$ is the same as FFJ with the relative minus sign. 

Whatever the momentum of the mother parton is, the NLO corrections to FFJ satisfies the sum rule: 
\be
\sum_l \int^1_0 dz z D_{J_l/k}^{(1)} (z) = 0,
\ee 
where again the superscript $(1)$ denotes the NLO correction.
Therefore the peturbative kernel $K_{m/k}$ satisfies the momentum conservation sum rule
\be
\sum_m \int^1_0 dz z K_{m/k} (z) = 1.
\ee


\section{Conclusions}
\label{conclude}

In this paper we introduce the fragmentation function  to a jet (FFJ), $D_{J_k/l}(z,\mu)$, which describes  the fragmentation of a parton $l$  into a jet with momentum fraction $z$ with parton $k$. This new object naturally appears in factorized rates when considering the jet radius, $R$, dependence.  To show this, we present a factorization theorem using SCET describing the rate for observing a fragmented hadron and a jet, which is the convolution of the partonic cross section, the FFJ, and the fragmentation of a hadron within a jet as shown in Eq.~(\ref{dxsec2}).

In order to resum the logarithms of $R$, we need the evolution equations for the FFJ.  We calculate the NLO corrections for all combinations of the quark and gluon-initiated to quark and gluon final state FFJs, and present the results in Eqs.~(\ref{Dqqj}), (\ref{Dgqj}), (\ref{gjetgfnlo}), and (\ref{gjetqfnlo}).
The one loop results of the FFJs satisfy the usual DGLAP evolution equations as seen in Eqs.~(\ref{pqq}-\ref{pgg}). This allows for the resummation of $\ln R$ using standard RG equation evolutions.

The formalism can be easily generalized to look at other interesting observables. For example, we  show how this formalism can be used to describe a subjet within a fat jet in Eq.~(\ref{dxsec3}).  This allows for the resummation of ratio of the radii of the jets. Using this improved theoretical prediction, we have a better theoretical description of this observable, which may be used to  investigate jet substructure as shown in Eqs.~(\ref{sJFFNLO}-\ref{sJFFqg}).  

As we will discuss in a forthcoming article \cite{piss}, resumming the $\ln R$ corrections can significant modify the cross sections.  In this follow-up paper, we will also show other places where the FFJ appears in theoretical predictions.  We will also look at the phenomenology of the subjet within a fat jet.

\begin{appendix}

\section{Hadron Fragmentation inside a Jet}
\label{A1}

We can describe the HFF inside a jet, $D_{H/J_k}(z)$, similar to sJFF. The unnormalized HFF inside a jet can be expressed as 
\bea\label{HFFjet} 
\tilde{D}_{H/J_q}(z,\mu) &=& \frac{z^{D-3}}{2N_c} \sum_{X {\in j}} \mr{Tr} \langle 0 | \delta \bigl(\frac{p_H^+}{z}-\mc{P}_+\bigr)\nn \Psi_n | H X\in J(p_J^+,R)\rangle \\
&&\times \langle HX \in J(p_J^+,R) | \bar{\Psi}_n |0\rangle. \nnb
\eea
Here we described the hardron fragmentation from the quark jet in the hadron frame $(\blpu{p}_H=0)$, and the momentum of the mother parton is given by $p_J$, hence $zp_J^+ = p_H^+$.

\begin{figure}[t]
\begin{center}
\includegraphics[height=5cm]{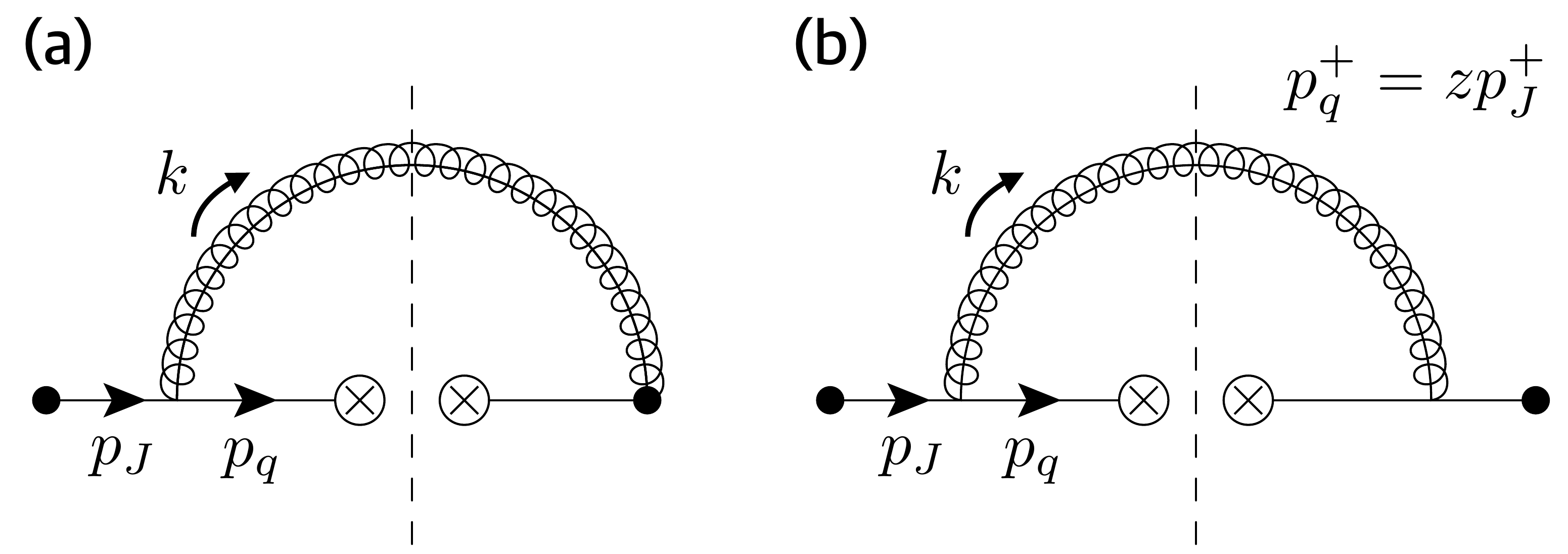}
\end{center}
\vspace{-0.3cm}
\caption{\label{fig5} \baselineskip 3.0ex 
Feynman diagrams of real gluon emissions for quark fragmentation inside a jet at NLO. The gluon in the final state is also inside a jet. Diagram (a) has its Hermitian conjugate contribution.
}
\end{figure}

Although the fragmentation function is a nonperturbative observable, it is important to understand its renormalization behavior computing the higher order corrections at the parton level separating IR divergences. At LO in $\as$, the fragmentation function from quark jet to quark is given by $D_{q/J_q}^{(0)}(z) = \delta(1-z)$. At NLO in $\as$, the virtual correction, including zero-bin subtraction,  is 
\be\label{qvirt} 
D_V = \frac{\as C_F}{\pi} \left(\frac{1}{\UV} - \frac{1}{\IR}\right)\left(\frac{1}{\UV}+\ln\frac{\mu}{p_J^+}+1\right) \delta(1-z).
\ee

The Feynman diagrams for real gluon emissions are shown in Fig.~\ref{fig5}, and only  diagram Fig.~\ref{fig5}-(a) has a nonvanishing zero-bin contribution. Thus the amplitude for Fig.~\ref{fig5}-(a) is written as 
\be\label{qMRa} 
D_R^{(a)} = \tilde{D}_R^{(a)} - D_{R,0}^{(a)}, 
\ee
where $\tilde{D}$ is the naive collinear contribution and $D_0$ is the zero-bin contribution. $\tilde{D}$ is given by 
\bea\label{qnaivea} 
\tilde{D}_R^{(a)}(z) &=& \frac{\as C_F}{2\pi} \frac{(\mu^2 e^{\gamma})^{\veps}}{\Gamma(1-\veps)} \int^{\La=p_J^{+2}t^2 z(1-z)}_{0} \frac{dM^2}{(M^2)^{1+\veps}} z^{1-\veps}(1-z)^{-1-\veps}\\
&=& \tilde{I}_{R}^{(a)} \delta(1-z) + \Bigl[D_R^{(a)}(z)\Bigr]_{+}\ , \nnb
\eea 
where $\tilde{I}_{R}^{(a)}$ is 
\bea 
\tilde{I}_{R}^{(a)} &=& \int^1_0 dz \tilde{D}_R^{(a)}(z) \nnb \\ 
\label{Iqa}
&=& \frac{\as C_F}{2\pi} \Biggl[\frac{1}{2\IR^2}+\frac{1}{\IR} \Bigl(1+\ln\frac{\mu}{p_J^+ t} \Bigr)+ 4-\frac{3\pi^2}{8} +\ln\frac{\mu}{p_J^{+2} t^2}+\frac{1}{4} \ln^2\frac{\mu}{p_J^{+2} t^2}\Biggr]\ . 
\eea

For the zero-bin contribution, the radiated gluon becomes soft and hence the $z$-dependence can be fixed as $\delta(1-z)$,  giving 
\bea 
D_{R,0}^{(a)}(z) &=& \frac{\as C_F}{2\pi} \frac{(\mu^2 e^{\gamma})^{\veps}}{\Gamma(1-\veps)} \delta(1-z) \int^{\infty}_0  dk_+ k_+^{-1-\veps} \int^{t^2k_+}_0 k_-^{-1-\veps} \nnb \\
\label{zbq}
&=& \frac{\as C_F}{2\pi} \Biggl[\frac{1}{2}\Bigl(\frac{1}{\UV}-\frac{1}{\IR}\Bigr)^2 +\Bigl(\frac{1}{\UV}-\frac{1}{\IR}\Bigr) \ln t \Biggr]\delta(1-z), 
\eea 
where the phase space constraint by the jet algorithm gives $t^2>k_-/k_+$  from Eqs.~(\ref{pscin}) and (\ref{pscout}), and the jet mass is approximated as $M^2 \sim p_J^+ k_-$. 

Similar to Eq.~(\ref{qnaivea}), the contribution of  diagram Fig~\ref{fig5}-(b) is
\bea\label{qb} 
D_R^{(b)}(z) &=& \frac{\as C_F}{2\pi} \frac{(\mu^2 e^{\gamma})^{\veps}}{\Gamma(1-\veps)} (1-\veps) \int^{\La=p_J^{+2}t^2 z(1-z)}_{0} \frac{dM^2}{(M^2)^{1+\veps}} z^{-\veps}(1-z)^{1-\veps}\\
&=& I_{R}^{(b)} \delta(1-z) + \Bigl[M_R^{(b)}(z)\Bigr]_{+}\ , \nnb
\eea 
where the integrated part $I_R^{(b)}$ is
\be\label{Iqb} 
I_{R}^{(b)}=-\frac{\as C_F}{2\pi} \Biggl[\frac{1}{2\IR}+\ln\frac{\mu}{p_J^+ t}+\frac{3}{2}\Biggr]\ .
\ee

Therefore combining Eqs.~(\ref{qvirt}), (\ref{Iqa}), (\ref{zbq}), and (\ref{Iqb}), we can obtain the part proportional to $\delta(1-z)$. This result should be equal to Eq.~(\ref{MinkT}), i.e., the integrated jet function at NLO for $\theta <R'$. This is confirmed by
\bea 
I_{q/q}^{\theta<R} \delta(1-z) &=& \mc{J}_q^{(1)} (\mu;E_JR') \delta(1-z) \nnb \\
&=& D_V + 2\bigl(\tilde{I}_R^{(a)} \delta(1-z)- M_{R,0}^{(a)}(z) \bigr) +I_R^{(b)} \delta(1-z) + \bigl(Z_{\xi}^{(1)} + R_{\xi}^{(1)} \bigr)\delta(1-z) \nnb \\
\label{IinR}
&=& \delta(1-z)\frac{\as C_F}{2\pi}\Biggl[\frac{1}{\UV^2}+\frac{1}{\UV}\Bigl(\frac{3}{2} +\ln\frac{\mu^2}{p_{J}^{+2}t^2}\Bigr) \\ 
&&~~~~~~~~~~~
+\frac{3}{2}\ln\frac{\mu^2}{p_{J}^{+2}t^2}+\frac{1}{2}\ln^2\frac{\mu^2}{p_{J}^{+2}t^2}+\frac{13}{2}-\frac{3\pi^2}{4} \Biggr]\ ,\nnb
\eea 
where $Z_{\xi}$ is the collinear quark field strength renormalization and $R_{\xi}$ is its residue. At one loop they are given by 
\be\label{qZR}
Z_{\xi}^{(1)} = -\frac{\as C_F}{4\pi} \frac{1}{\UV}\ ,~~~ R_{\xi}^{(1)} = \frac{\as C_F}{4\pi} \frac{1}{\IR}\ .
\ee

The remaining distribution parts in Eqs.~(\ref{qnaivea}) and (\ref{qb}) are 
\bea
\Bigl[D_R(z)\Bigr]_+ &=& \Bigl[2D_R^{(a)}(z)+D_R^{(b)}(z)\Bigr]_+ \nnb \\
\label{qdist1}
&=&-\frac{\as C_F}{2\pi} \Biggl[\frac{1+z^2}{1-z} \Bigl(\frac{1}{\IR}+\ln\frac{\mu^2}{p_J^{+2}t^2} - 2\ln z(1-z) \Bigr)-(1-z)\Biggr]_+ \\
\label{qdist2}
&=&-\frac{\as C_F}{2\pi}\Biggl\{\delta(1-z)\Biggl[\frac{3}{2}\Bigl(\frac{1}{\IR}+\ln\frac{\mu^2}{p_J^{+2}t^2}\Bigr)+\frac{13}{2}-\frac{2\pi^2}{3}\Biggr] \\
&&+(1+z^2)\Biggl[\frac{1}{(1-z)_+}\Bigl(\frac{1}{\IR}+\ln\frac{\mu^2}{p_J^{+2}t^2}-2\ln z\Bigr)-2\left(\frac{\ln(1-z)}{1-z}\right)_+\Biggr]-(1-z)\Biggr\}\ {.} \nnb
\eea

Finally, combining Eqs.~(\ref{IinR}) and (\ref{qdist2}), we obtain the unnormalized HFF inside a jet up to NLO,
\bea \label{HFFJNLO}
&&\tilde{D}_{q/J_q}(z,\mu;E_JR') = \mc{J}_q (\mu;E_JR') \delta(1-z) +  \Bigl[D_R(z)\Bigr]_+ \\
&&~~~~~ = \delta (1-z) \Biggl\{1+\frac{\as C_F}{2\pi}\Biggl[\frac{1}{\UV^2}+\frac{1}{\UV}\ln\frac{\mu^2}{p_{J}^{+2}t^2}+\frac{3}{2}\Bigl(\frac{1}{\UV}-\frac{1}{\IR}\Bigr) 
+\frac{1}{2}\ln^2\frac{\mu^2}{p_{J}^{+2}t^2}-\frac{\pi^2}{12} \Biggr]\Biggr\}\nnb\\ 
&&~~~~~~
-\frac{\as C_F}{2\pi}\Biggl\{(1+z^2)\Biggl[\frac{1}{(1-z)_+}\Bigl(\frac{1}{\IR}+\ln\frac{\mu^2}{p_J^{+2}t^2}-2\ln z\Bigr)-2\left(\frac{\ln(1-z)}{1-z}\right)_+\Biggr]-(1-z)\Biggr\}\ {.} \nnb
\eea
The normalized HFF inside a jet  is obtained by dividing by $\mc{J}_q(\mu;E_JR')$,
\bea
D_{q/J_q}(z) &=& \frac{\tilde{D}_{q/J_q}(z;E_JR')}{\mc{J}_q(\mu;E_JR')} \nnb\\
\label{nHFFJNLO}
&=& \delta(1-z) -\frac{\as}{2\pi}\Biggl\{P_{qq} (z) \Bigl(\frac{1}{\IR}+\ln \frac{\mu^2}{p_J^{+2}t^2}\Bigr) 
+ C_F \Biggl[\delta(1-z) \Bigl(\frac{13}{2}-\frac{2\pi^2}{3}\Bigr)-(1-z) \nnb \\
&&-2(1+z^2)\left(\frac{\ln z}{(1-z)_+}+\Bigl(\frac{\ln(1-z)}{1-z}\Bigr)_+\right)\Biggr]\Biggr\}\ .
\eea

In a similar way we can compute the other HFFs inside a jet. Their NLO results are
\bea
\label{nHFFgq}
D_{g/J_q}(z) &=& -\frac{\as}{2\pi} \Biggl[P_{gq}(z) \Bigl(\frac{1}{\IR}+\ln\frac{\mu^2}{p_J^{+2}t^2}  - 2\ln z(1-z)\Bigr)-z C_F\Biggr]\ ,\\
\label{nHFFgg}
D_{g/J_g}(z) &=& \delta(1-z) -\frac{\as}{2\pi}\Biggl\{P_{gg} (z) \Bigl(\frac{1}{\IR}+\ln \frac{\mu^2}{p_J^{+2}t^2}\Bigr)  +N_c\Biggl[\delta(1-z)\Bigl(\frac{67}{9}-\frac{23n_f}{18N_c}-\frac{2\pi^2}{3}\Bigr)\nnb\\
&&\hspace{-1cm}-4\Bigl[\frac{z\ln z}{(1-z)_+}+z\Bigl(\frac{\ln(1-z)}{1-z}\Bigr)_+ + \ln [z(1-z)]\Bigl(\frac{1-z}{z}+z(1-z)\Bigr)\Bigr]\Biggr\}\ ,\\
\label{nHFFqg}
D_{q/J_g}(z) &=& -\frac{\as}{2\pi} \Bigl[P_{qg}(z) \Bigl(\frac{1}{\IR}+\ln\frac{\mu^2}{p_J^{+2}t^2}-2\ln [z(1-z)]\Bigr)-z(1-z)\Bigr]\ .
\eea

At much lower energy scale, $\mu \ll p_J^+ t$, the fragmenting process cannot resolve the scale $p_J^+ t$. Hence 
the scale $p_J^+ t$ can be identified as an UV scale. In this case the fragmenting process can be described by the standard fragmentation function without the  phase space restriction. 
Therefore, similar to the subjet case shown in Eq.~(\ref{facD3}), the FF inside a jet is in general factorized as follows~\cite{Procura:2011aq}: 
\be\label{facD1} 
D_{l/J_k}(z,\mu;E_JR') = \int^1_z \frac{dx}{x} K_{m/k} (z/x,\mu;E_JR') D_{l/m} (x,\mu),
\ee 
where $k,~l$, and $m$ represent the quark flavors and gluon, and $m$ is the dummy index. If we consider the HFF, we have  
\be\label{facD2} 
D_{H/J_k}(z,\mu;E_JR') = \int^1_z \frac{dx}{x} K_{m/k} (z/x,\mu;E_JR') D_{H/m} (x,\mu).
\ee 
Here $K_{m/k}$ are the perturbative kernels with a typical energy scale $p_J^+ t\sim E_JR'$. They are obtained from the matching between two fragmentation functions. Because $K_{m/k}$ is irrelevant to the lower energy scale dynamics, they are universally given when we consider a fragmentation process inside a jet. 

Under dimensional regularization, the bare result of NLO corrections to the standard fragmentation function at parton level is
\be 
\label{FFnlo}
D_{l/m}^{(1)}(z) = \frac{\as}{2\pi} P_{lm}(z) \Bigl(\frac{1}{\UV}-\frac{1}{\IR}\Bigr)\ ,
\ee
where $P_{lm}$ are DGLAP splitting kernels. Comparing the NLO results of the HFF inside a jet and Eq.~(\ref{FFnlo}), we can easily check that the kernels in Eqs.~(\ref{facD1}) and (\ref{facD2}) are the same as ones for the subjet case in sec.~\ref{subjet}.

\end{appendix}

\acknowledgments

C.~K. was supported by Basic Science Research Program through the National Research Foundation of Korea (NRF) funded by the Ministry of Science, ICT, and Future Planning (Grant No. NRF-2014R1A2A1A11052687). A.~L. and L.~D. were supported in part by NSF Grant No. PHY-1519175.



\end{document}